\documentclass[pra,aps,twocolumn,showpacs,showkeys]{revtex4}
\usepackage{graphicx}
\usepackage{amsmath}
\newcommand{\ket}[1]{|#1\rangle}
\newcommand{\mket}[1]{$|#1\rangle$}
\newcommand{\bra}[1]{\langle#1|}

\newcommand{\be}{\begin{equation}}
\newcommand{\eel}[1]{\label{#1} \end{equation}}
\newcommand{\ee}{\end{equation}}
\newcommand{\al}{\alpha}

\begin{document}
\title{Methods for Producing Optical Coherent State Superpositions}
\author{S. Glancy}
\affiliation{Mathematical and Computational Sciences Division, National Institute of Standards and Technology, Boulder, Colorado 80305, USA}
\email{sglancy@boulder.nist.gov}
\author{H. M. Vasconcelos}
\email{hilma@ufc.br} 
\affiliation{Departamento de Engenharia Metal\'ugica e de Materiais, Universidade Federal do Cear\'a,
Campus do Pici - Bloco 714, Fortaleza, Cear\'a 60455-760, Brazil}
\date{2008 April 28}
\begin{abstract}

We discuss several methods to produce superpositions of optical coherent states (also known as ``cat states'').  Cat states have remarkable properties that could allow them to be powerful tools for quantum information processing and metrology.  A number of proposals for how one can produce cat states have appeared in the literature in recent years.  We describe these proposals and present new simulation and analysis of them incorporating practical issues such as photon loss, detector inefficiency, and limited strength of nonlinear interactions.  We also examine how each would perform in a realistic experiment.
\end{abstract}
% \ocis{270.0270, 270.5585, 270.6570,190.3270}
\maketitle

\section{\label{Introduction}Introduction}

For many years a primary goal for researchers in quantum optics has been the generation of exotic quantum states of light.  These have 
included squeezed vacuum states, photon number eigenstates (also known as Fock states), and many examples of entangled sets of 
photons.  States such as these have served well in many basic tests of the foundations of quantum theory, and they may eventually prove to 
be useful for quantum computation, quantum communication, quantum metrology, and lithography. \cite{bachor2004} Therefore there is much interest in 
improving methods to produce these states and to generate other types of optical quantum states.         

The particular quantum state of interest in this paper is a superposition of two coherent states with opposite phase, which is often referred 
to as a (Schr\"odinger) cat state.  Cat states may be used as the logical qubit basis in a quantum computer \cite{jeong2001, ralph2003}.  They may also serve as input states to an interferometer that is able to measure distances with greater accuracy than achievable within the 
limits usually imposed by the light's wavelength \cite{ralph2001}.  Transforming a single coherent state into a cat state through unitary 
evolution alone would require a strong nonlinearity.  Also, cat states are extremely sensitive to decoherence from photon absorption.  For 
these reasons cat states containing more than one photon on average have been produced only in cavity-quantum-electrodynamic experiments in which an atom interacts with the electromagnetic field confined to a high finesse optical cavity \cite{brune1996,auffeves2003}.  In experiments of this sort the cavity confines the optical mode to a small volume so that it interacts very strongly with an atom passing through the cavity.  Unfortunately, because the cat state is confined to a cavity, it can neither be manipulated with tools such as beam splitters or phase shifters, nor be measured with standard optical means 
such as photon counters or homodyne detection.  For the uses described in \cite{jeong2001, ralph2003, ralph2001} we require cat states that occupy freely propagating optical modes.

In recent years researchers have proposed several schemes to produce freely propagating cat states.  The purpose of this paper is to provide a comprehensive review 
and critique of these schemes.  We examine the performance of these schemes in realistic experimental environments subject to problems such as photon loss, detector inefficiency and noise, and limited strength of nonlinear interactions.  We also make recommendations for the design of cat production experiments.  In this introductory section we review the properties of cat states and 
optical tools such as beam splitters and photon counters, which are commonly used in the cat production schemes.  The introduction also describes our model for photon absorption, how photon absorption affects cat states, and the difficulties of verifying that a cat state has 
been produced in an experiment.  Section 2 examines a method originally proposed by Yurke and Stoler \cite{yurke1986} to transform a 
coherent state into a cat state using the optical Kerr effect.  Section 3 briefly discusses the suggestion by Wolinsky and Carmichael \cite{wolinsky1988} that one may make cat states using a degenerate parametric oscillator in the strong coupling regime.  Section 4 examines the method proposed by Song, Caves and Yurke 
\cite{song1990} to produce cat states based on an optical backaction evasion measurement.  Section 5 examines the method of photon 
subtraction from a squeezed vacuum state proposed by Dakna and co-authors \cite{dakna1997}.  Section 6 examines a method from Lund, Jeong, 
Ralph, and Kim \cite{lund2004} by which one may use small amplitude cat states, linear optical devices, and measurements to produce large 
amplitude cats.  Section 7 examines two schemes \cite{jeong2004,gerry1999} that use a weak Kerr effect followed by measurements to make high amplitude 
cats.  We conclude with some general observations in Section 8.

A few experiments have already demonstrated the production of small cat states containing less than one photon \cite{wenger2004b,neergaard-nielsen2006,ourjoumtsev2006,wakui2006}.  These are sometimes called ``Schr\"odinger kittens'' or simply ``non-Gaussian states''.  All of these experiments use the photon subtraction method and will be discussed in Section 4.

\subsection{Properties of Cats}  For a general introduction to the quantum properties of light, we refer the reader to the text \cite{mandel1995}.  The electromagnetic field may be decomposed into independent modes, each of which is defined by a particular polarization, distribution of light frequencies, 
and direction of propagation.  Each mode is modeled as a quantum mechanical oscillator whose frequency is equal to the light's frequency 
$\omega$.  The number of photons occupying a mode is equal to the number of quantized energy excitations in that mode's oscillator.  Let 
$\hat{a}$ be the photon annihilation operator, so $\hat{a}^\dagger$ is the creation operator, and $\hat{a}^\dagger \hat{a} = \hat{n}$ is 
the operator corresponding to the photon number observable.  We use $|n\rangle$ as the photon number eigenstate, so 
$\hat{n}|n\rangle=n|n\rangle$.  For cases in which we are interested in more than one mode, we will distinguish the operators and 
states with subscripts labeling each mode.

The coherent state $|\alpha\rangle$ is the eigenstate of the annihilation operator with eigenvalue $\alpha$:
\begin{equation}
\hat{a}|\alpha\rangle=\alpha|\alpha\rangle,
\end{equation}
where $\alpha$ may be any complex number.  The light produced by ordinary lasers well approximates a coherent state. (In fact it is more accurate to say that laser light approximates the mixed state $\int d\phi |e^{i\phi}\alpha\rangle\langle e^{i\phi}\alpha |$, but if all light modes are phase locked to a local oscillator, then this distinction is irrelevant.) In the photon number basis a coherent state has the decomposition
\begin{equation}
|\alpha\rangle=e^{-|\alpha|^2/2}\sum_{n=0}^{\infty}\frac{\alpha^n}{\sqrt{n!}}|n\rangle.
\end{equation}
For ease of notation we usually assume $\alpha$ is a real, positive number.  In a few cases we will multiply $\alpha$ by a complex phase to characterize coherent states with complex amplitudes.

We define a cat state to be a superposition of two coherent states with opposite phases:
\begin{equation}
|\Psi_{\pm}(\alpha)\rangle=\frac{1}{\sqrt{N_{\pm}(\alpha)}}(|-\alpha\rangle\pm|\alpha\rangle).
\label{cat_state_definition}
\end{equation}
The normalization factor $N_{\pm}(\alpha)=2\pm 2e^{-2\alpha^2}$ is required because the states $|-\alpha\rangle$ and 
$|\alpha\rangle$ are not orthogonal.  However $|\langle-\alpha|\alpha\rangle|^2=e^{-4\alpha^2}$, so that for a moderately sized $\alpha=2$ this overlap is only 
$1.1 \times 10^{-7}$.  When expressed in the photon number basis $|\Psi_{\pm}(\alpha)\rangle$ becomes
\begin{equation}
|\Psi_{\pm}(\alpha)\rangle=\frac{e^{-\alpha^2/2}}{\sqrt{N_{\pm}(\alpha)}} \sum_{n=0}^{\infty} \frac{\alpha^n}{\sqrt{n!}}\left( (-1)^n\pm 1 
\right)|n\rangle.
\end{equation}
$|\Psi_{+}(\alpha)\rangle$ contains only even numbers of photons, and $|\Psi_{-}(\alpha)\rangle$ contains only odd numbers.  For this reason they are often called the ``even cat'' and ``odd cat'' states.  The even and odd cat states are orthogonal to one another, and one may perform a measurement to distinguish the two by 
counting the photons.

In addition to the photon number basis we may also measure the cat state in the quadrature bases corresponding to the position and momentum 
of the oscillator.  The position observable is defined by
\begin{equation}
\hat{x}=\frac{1}{\sqrt{2}}\left( \hat{a}^\dagger +\hat{a}\right),
\end{equation}
and the momentum observable is
\begin{equation}
\hat{p}=\frac{i}{\sqrt{2}}\left( \hat{a}^\dagger -\hat{a}\right).
\end{equation}
In the $\hat{x}$ basis, the cat states $|\Psi_\pm(\alpha)\rangle$ have the wave functions
\begin{equation}
\Psi_\pm^{(\alpha)}(x)=\frac{\pi^{-1/4}}{\sqrt{N_\pm(\alpha)}}\left( e^{-\frac{1}{2}\left( x+\sqrt{2}\alpha\right)^2} \pm e^{-\frac{1}{2}
\left( x-\sqrt{2}\alpha\right)^2}\right).
\label{cat_x_wavefunction}
\end{equation}
It has two characteristic Gaussian shaped humps located at $x=\pm \sqrt{2}\alpha$.  In the $\hat{p}$ basis the cat states have the wave 
functions
\begin{eqnarray}
\Psi_+^{(\alpha)}(p) & = & \frac{2\pi^{-1/4}}{\sqrt{N_\pm(\alpha)}}e^{-p^2/2}\cos(\sqrt{2}p\alpha) \\
\Psi_-^{(\alpha)}(p) & = & \frac{2\pi^{-1/4}}{\sqrt{N_\pm(\alpha)}}e^{-p^2/2}i\sin(\sqrt{2}p\alpha). 
\end{eqnarray}
These are characterized by oscillations of frequency $\sqrt{2}\alpha$ inside a Gaussian envelope centered at $p=0$.  Resolving these 
oscillations is another possible method to distinguish between even and odd cats.

We will also consider slightly more general cats with phase $\phi$:
\be
\ket{\Psi_\phi(\alpha)}=\frac{1}{\sqrt{N_\phi(\alpha)}}\left(\ket{-\alpha}+e^{i\phi}\ket{\alpha}\right).
\ee
$\ket{\Psi_0(\alpha)}=\ket{\Psi_+(\alpha)}$ and $\ket{\Psi_\pi(\alpha)}=\ket{\Psi_-(\alpha)}$.  Changing $\phi$ changes the number of photons contained in $\ket{\Psi_\phi(\alpha)}$ and requires some type of nonlinear interaction.

\subsection{Beam Splitters} Many of the following cat state production schemes use beam splitters, so we will introduce them here. For a 
more thorough discussion of beam splitters see \cite{leonhardt1997}.  One may think of a beam splitter as a partly silvered mirror.  A mode of light enters the beam splitter on each of the mirror's two faces.  Some fraction $T$ of the light energy in each mode is 
transmitted through the mirror and $R=1-T$ is reflected.  While passing through the beam splitter, the state of the two input modes will 
evolve according to the transformation
\begin{equation}
|\psi\rangle_1|\phi\rangle_2 \rightarrow \hat{B}(T)|\psi\rangle_1|\phi\rangle_2,
\end{equation}
where $\hat{B}(T)$ is the unitary operator given by
\begin{equation}
\hat{B}(T)=e^{\arccos(\sqrt{T})(\hat{a}_1\hat{a}_2^\dagger-\hat{a}_1^\dagger\hat{a}_2)}.
\end{equation} 
When two coherent states $|\alpha\rangle_1$ and $|\beta\rangle_2$ enter a beam splitter, their state will become
\begin{equation}
|\alpha\rangle_1|\beta\rangle_2 \rightarrow \left|\alpha\sqrt{T}-\beta\sqrt{1-T}\right\rangle_1 \left|\beta\sqrt{T}+\alpha\sqrt{1-T}\right\rangle_2.
\end{equation}
It should be clear from this relation that using only coherent states and beam splitters one can never produce a cat state, because a beam 
splitter simply transforms coherent states into coherent states of different amplitudes.

\subsection{Photon Counters}  To make cats we will imagine that we have devices that can count the number of photons in an optical mode.  
Such devices have been very difficult to engineer, but 
significant progress has been made in this area during recent years.  Photon counters suffer from inefficiency and dark counts.  Inefficient detectors fail to register all photons arriving in the mode of interest.
%This may be caused by reflection from the detector's face, the photon traveling through the detector without interaction, or the photon begin absorbed by some part of the detector which does not register the photon's presence.
We model this by imagining that the mode of interest passes through a beam splitter whose transmissivity is equal to the detector's efficiency $\eta$ before entering an ideal detector \cite{leonhardt1997}.  If the signal mode has a probability distribution to measure 
$n$ photons given by ${\rm P}(n)$, then the detector will have a probability of ${\rm P}_\eta(m)$ to register $m$ photons:
\begin{equation}
{\rm P}_\eta(m)=\sum^{\infty}_{n=m} \binom{n}{m} \eta^m (1-\eta)^{n-m} {\rm P}(n),
\label{inefficiency_transformation}
\end{equation}
when $m \leq n,$ and ${\rm P}_\eta(m)=0$ when $m>n$.

In addition to inefficiencies, detectors also suffer from dark counts.  These are events in which the detector registers more photons than in fact arrive in the mode we are measuring.
%Dark counts may be caused by for example thermal excitations of electrons in the detector and ambient light entering the detector's face through the signal mode or some other mode.
To model the dark counts we imagine that a second 
light mode (in addition to the mode we are measuring) enters the detector.  We assume the number of photons in this second mode has a Poisson 
probability distribution.  If the mean number of dark counts is $d$ during a detection event of fixed duration, then the probability that $q$ dark count photons are registered is 
\begin{equation}
{\rm p}_{d}(q)=\frac{d^q e^{-d}}{q!}.
\label{dark_count_probability}
\end{equation}
Given an initial photon distribution in the signal mode of ${\rm P}(n)$, we combine the dark count and inefficiency influences to find 
that the probability to register $m$ photons is
\begin{eqnarray}
{\rm P}_d(m) & = & \sum_{x=0}^{m}{\rm P}_{\eta}(x){\rm p}_d(m-x) \\
 & = & \sum_{x=0}^{m}{\rm p}_d(m-x) \nonumber \\
 & & \times \sum^{\infty}_{n=x} \binom{n}{x} \eta^x (1-\eta)^{n-x} {\rm P}(n).
\label{dark_count_transformation}
\end{eqnarray}
Here the sum adds the probabilities that $x$ photons are transmitted through the beam splitter and that $m-x$ additional dark counts occur for all $x \leq m$.

Photon detectors are usually characterized by their dark count rate $D$, the number of dark counts registered per unit time.  The size of $d$ 
will depend on each detector's dark count rate and the duration of the observation.  Shorter observing times will give a smaller $d$ but require faster detectors.  Our model for photon counters does not incorporate the effects of dead time.  We assume that such effects can be made to be negligible by increasing the time between detection events.

Let us briefly describe two photon counters that may be suitable to assist in a cat production experiment: the visible light photon 
counter (VLPC) and the superconducting transition edge sensitive photon counter (TESPC). The VLPC is discussed in \cite{takeuchi1999}.  
It is essentially an array of avalanche photodiode detectors (APDs), which operate by a version of the photoelectric effect in which incident light 
frees an electron from the valence band lifting it into the conducting band.  A large voltage then accelerates this electron, which collides 
with others, creating an avalanche of electrons that is amplified and manipulated with electronic devices.  Most avalanche photodiodes are 
unable to distinguish between 1, 2, 3, ... photons.  Because a single photon depletes the available electrons, the arrival of a second photon 
is unable to create a larger electronic signal. The VLPCs can distinguish between different numbers of photons because the avalanche is 
confined to a small portion of the detector's area, leaving most of the VLPC's area available for the detection of the other photons.  
References \cite{kim1999,waks2003a,waks2003b,waks2003c} discuss the use of the VLPC to distinguish between $m-1$ and $m$ photons with $m$ as large 
as 10 photons.  They also successfully demonstrate the use of the transformation Eq. (\ref{inefficiency_transformation}) to correct for 
detector inefficiencies.  In \cite{takeuchi1999} Takeuchi and co-authors report achieving a maximum single photon quantum efficiency of 
$\eta=0.882 \pm 0.05$ using 694 nm light, with a dark count rate of $D=2 \times 10^4$ counts/second. Lower dark count rates may be achieved 
by reducing bias voltage applied to the photodiodes, but this also reduces the efficiency.  They also claim that their photon counting system roughly obeys the relation
\begin{equation}
\log_{10}(D)=3.226 \eta + 1.206,
\end{equation}
where $D$ is measured in counts per second.  In \cite{waks2003a}, Waks and co-authors describe an experiment in which the VLPC was used to 
observe the photon statistics of a squeezed vacuum state.  They produced squeezed light of 532 nm pulsed at 20 ns.  By limiting the detection 
integration window to 20 ns they were able to reduce the average number of dark counts per detection event to $d \sim 4\times 10^{-4}$.

The TESPC detects photons through calorimetry \cite{rosenberg2005}.  The light is absorbed by the superconducting 
electrons in a small sample of tungsten.  The tungsten sample is voltage biased on the edge of the superconducting-to-normal transition, so a small increase in the electrons' temperature creates a large change in the tungsten's resistance.  Measuring this resistance change allows one to calculate the energy deposited by the incoming photons.  Rosenberg and co-authors \cite{rosenberg2005} report an efficiency of 
$\eta=0.88$ at a wavelength of 1550 nm.  They observed a dark count rate of only $D=400$ counts/second.  This is caused primarily by blackbody
radiation entering the optical fiber delivering the photons to the detector and ambient light scattering into the fiber.  Therefore the 
dark count rate should be significantly reduced for light at shorter wavelengths.  Because these primary dark count sources are not intrinsic 
to the TESPC, it is likely that the dark count rate can be reduced to negligible levels with sufficient technical care (for example, shielding 
and filtering) of the light path.  The recent advent of such high efficiency, low dark count detectors such as the VLPC and the TESPC makes 
several of the following cat production methods just now feasible. 

\subsection{Cat State Decoherence.} In addition to the difficulty of creating cat states we must also face the extreme fragility of these states to 
decoherence.  We expect that the primary source of decoherence in an optical experiment will be photon absorption.  This can be modeled 
by assuming some of the cat state is lost via a beam splitter type of interaction.  The cat state enters one mode of the beam splitter, and 
the vacuum enters the other mode.  After passing through the beam splitter some of the cat's energy will be transferred to the second mode 
and lost to the environment.
%Only a single beam splitter is necessary to characterize the effects of any number of photon loss events and mechanisms.  Even though the photons will actually be lost to a myriad of different modes, our assumption that all photons are lost to a single mode is sufficient, because we must trace over this mode (or these modes) to calculate the state of the transmitted qubit \cite{leonhardt1997}.
Suppose the cat state passes through some medium whose transmissivity is 
$\eta$, the cat occupies mode 1, and mode 2 contains the vacuum used to model the environment.  The initial state of the cat/environment 
system is $|\Psi_{\pm}(\alpha)\rangle_1|0\rangle_2$.  After the two modes pass through the beam splitter their state becomes
\begin{equation}
|-\alpha \sqrt{\eta}\rangle_1|-\alpha \sqrt{1-\eta}\rangle_2 + |\alpha \sqrt{\eta}\rangle_1|\alpha \sqrt{1-\eta}\rangle_2,
\end{equation}
where we have omitted the normalization.  The second mode is lost to the environment, and the first mode is in a mixed state given by
\begin{eqnarray}
\rho_\pm(\alpha,\eta) & = & 
(1-P_\pm)
|\Psi_\pm(\alpha\sqrt{\eta})\rangle\langle\Psi_\pm(\alpha\sqrt{\eta})|\nonumber \\
 & & +
P_\pm
|\Psi_\mp[(\alpha\sqrt{\eta})\rangle\langle\Psi_\mp(\alpha\sqrt{\eta})|,
\end{eqnarray}
where
\begin{equation}
P_\pm=\frac{1}{2}\frac{N_\mp(\alpha\sqrt{\eta})}{N_\pm(\alpha)} \left( 1-e^{-2\alpha^2(1-\eta)}\right),
\end{equation}
and $N_\pm$ is defined just following Eq.~(\ref{cat_state_definition}).  Absorption affects the cat states
in two ways.  The amplitude of the coherent states is decreased from $\alpha$ to $\alpha\sqrt{\eta}$.  With a probability $P_+$ the even cat 
becomes an odd cat, and with probability $P_-$ the odd cat becomes an even cat.  These probabilities increase quickly with increasing $\alpha$.  This result is discussed in more detail in \cite{cochrane1998, glancy2003}.

\begin{figure}
\includegraphics[width=8.255cm]{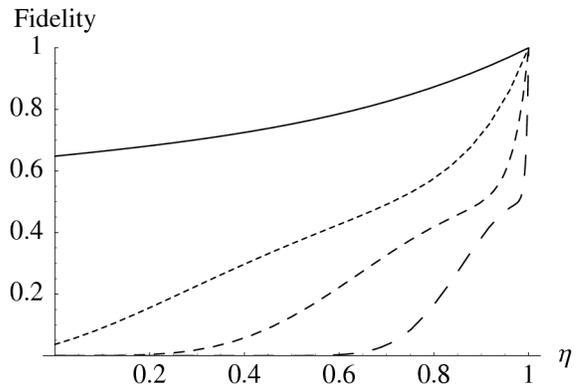}
\caption{Plot of $F_+(\alpha,\eta)$, the fidelity of a cat state $|\Psi_+(\alpha)\rangle$ that has suffered from some decoherence by passing through a medium of transmissivity $\eta$. The plot includes curves for even cats with $\alpha=1$ (solid curve), $\alpha=2$ (small dashes), $\alpha=4$ (medium dashes), $\alpha=10$ (large dashes).
\label{cat_decoherence}}
\end{figure}

To compare the decohered cat state with the original cat state, we will use the fidelity, calculated according to
\begin{equation}
F_\pm(\alpha,\eta)=\langle \Psi_\pm(\alpha)|\rho_\pm(\alpha,\eta)|\Psi_\pm(\alpha)\rangle.
\end{equation}
We find that
\begin{equation}
F_\pm(\alpha,\eta)=(1-P_\pm)\frac{4e^{-\alpha^2(1+\eta)}}{N_\pm(\alpha)N_\pm(\alpha\sqrt{\eta})}\left( e^{\alpha^2\sqrt{\eta}}\pm e^{-\alpha^2
\sqrt{\eta}}\right)^2.
\end{equation}
In Fig.~\ref{cat_decoherence} we investigate the dependence of this fidelity on $\alpha$ and $\eta$.  This plot shows a very large loss of fidelity for higher amplitude cat states even for $\eta$ very close to 1.  We also see the distinct behavior of the two types of error.  For large
$\alpha$ the fidelity quickly drops to $1/2$ as $\eta$ decreases from 1, because the initial pure cat becomes the mixed state 
$1/2 |\Psi_+(\alpha)\rangle \langle \Psi_+(\alpha)|
 + 1/2|\Psi_-(\alpha)\rangle \langle \Psi_-(\alpha)|$.  As $\eta$ continues to decrease, the coherent state amplitude decreases, lowering the fidelity from $1/2$ to 0.

\subsection{Cat State Verification.} Any cat generation experiment requires a verification that the experiment  makes cat states.  This is likely to be accomplished by performing a series of homodyne measurements, which can be used to reconstruct the state through quantum tomography \cite{leonhardt1997,lvovsky2005}.  If the photodetectors used in the homodyne measurement have efficiency 
$\eta$, then the homodyne measurement signal will be equivalent to that of the generated state having passed through a beam splitter with transmissivity $\eta$.  If  $\eta$ is known, the homodyne detection inefficiency effects can be corrected in data analysis by using the inverse Bernoulli transformation (based on Eq.~(\ref{inefficiency_transformation})) and other techniques described in \cite{leonhardt1997,lvovsky2005}.  The correction is approximate and depends on the accuracy with which the efficiency is known, what other error sources are present, and the algorithm used to infer the quantum state from the homodyne data.  Fortunately, the photodetectors used for homodyne measurements have much higher efficiencies than those used for counting small numbers of 
photons.  For example Polzik, Carri, and Kimble report obtaining an overall homodyne detection efficiency of $\eta=0.98$ in \cite{polzik1992}.

Reconstructing a cat state through tomography requires a large number of homodyne measurements.  In this procedure one measures the probability distribution of the $\hat{x}$ quadrature for each phase of the local oscillator from 0 to $\pi$.  In practice this is done 
by constructing a histogram of $\hat{x}$ with a finite bin size, with a finite number of measurements, for a finite number of local oscillator
phases.  We would like to estimate the required number of phases, the bin size resolution, and the number of measurements required for each 
histogram.  Determining the best method for assigning uncertainty to, for example, the density matrix elements of the reconstructed state and the relationship between error bars and the number of measurements requires further research.  For discussion of these issues see \cite{paris2004,lvovsky2005,blume-kohout2006}.  A full treatment of this difficult data analysis problem requires further research.  Instead, we quote a few rules 
of thumb from \cite{leonhardt1997}, which contains a much more detailed discussion.  To reconstruct the state of a mode 
containing at most $M$ photons, one should collect histograms for $M+1$ equally spaced local oscillator phases, and the histogram bin size 
resolution must be less than $\pi/(2\sqrt{2M+1})$.  Suppose the state we are measuring has the photon number probability distribution 
${\rm P}(m)$.  To accurately reconstruct the probability to detect $m$ photons each histogram should contain on the order of 
$4\times ({\rm P}(m))^{-2}$ counts.  These rules can help us to estimate the order of magnitude of the number of individual homodyne 
measurements required to verify the production of a cat state.  For example, the even $\alpha=2$ cat contains a mean number of approximately 
4 photons.  The probability to detect 10 photons in this cat state is ${\rm P}(10)\sim 0.01$.  If we are willing to ignore higher photon 
numbers, reconstructing the Wigner function of this state will require on the order of $4\times (0.01)^{-2}\times (10+1) \sim 4\times 10^6$ 
individual homodyne measurements.  For the more modest goal of verifying an $\alpha=1$ cat, we could ignore photon numbers larger than 4.  
In this case ${\rm P}(4)=0.03$, so we would need $\sim 2 \times 10^5$ measurements.  These rough estimates highlight the need for any proposed experiment to be able to produce a large number of cat states in order to perform a full reconstruction of the state generated by that experiment.

It is possible to estimate the fidelity of the state produced in an experiment with a perfect cat state without fully reconstructing the unknown state.  See \cite{somma2006} for an application of such a technique in an ion trap system.

\section{\label{section:kerr}Kerr Effect}

In 1986, Yurke and Stoler \cite{yurke1986} considered the evolution of a coherent state under the influence of the Kerr effect. They showed that, under suitable conditions, the coherent state evolves into a cat state.

The Kerr effect causes a phenomenon known as ``self-phase modulation.''  When light passes through a medium exhibiting the Kerr effect, the rate at which the light's phase advances depends on the intensity of the light.  This is modeled with an anharmonic-oscillator Hamiltonian of the form
\begin{equation}
H_{K} = \omega\hat{n}+\chi \hat{n}^2,
\label{Kerr-Hamiltonian}
\end{equation}
where $\omega$ is the light frequency, and $\chi$ is the strength of the nonlinear term.  The first term causes only linear phase evolution, so we neglect it in the following treatment.

If we apply this Hamiltonian to the coherent state $\ket{\beta}$ for a time $\frac{\pi}{2 \chi}$,
\be
\ket{\beta}\rightarrow \ket{\Psi_{\frac{\pi}{2}}(\beta)}=\frac{1}{\sqrt{2}}\left(\ket{-\beta}+i\ket{\beta}\right).
\label{-icat}
\ee
Due to the $i$ factor, this cat is slightly different from those discussed in other sections of this paper, but the $i$ cat exhibits the same interesting and useful features.  Achieving this transformation requires a large nonlinear strength $\chi$ or a long interaction time.

The Hamiltonian in Eq. (\ref{Kerr-Hamiltonian}) is naive and cannot capture the full dynamics of the light's interaction with the nonlinear medium.  We assigned a single frequency to the light, and only continuous waves of infinite duration can have a single frequency.
%If a continuous wave passes through some chunk of nonlinear medium, our treatment should account for changes to the light mode on entrance and exit from the medium.
If a pulse of light with a finite duration, with say a Gaussian shape, interacts with the nonlinear medium the dynamics will be far more complicated.  Because the pulse will have lower intensity (or photon number density) at its leading and trailing edges, the Kerr effect will have a greater effect on the center of the pulse.  Any linear or nonlinear dispersive properties of the medium will distort the pulse shape.  We are ignoring all of these effects in the following treatment.  Instead, our goal is to provide the first analysis of the relationship between cat state fidelity, $\chi$, and absorption.

To model loss we assume that the number of photons decays in time according to
\be
\langle \hat{n}\left(t\right)\rangle = \langle\hat{n}\left(0\right)\rangle e^{-\gamma t},
\ee
where $\gamma$ describes each material's photon loss rate.  The evolution of a (possibly mixed) state traveling through a Kerr medium with loss obeys the master equation
\be
\frac{\mathrm{d}\hat{\rho}}{\mathrm{d}t} = -i\chi\left[\hat{n}^2,\hat{\rho}\right] + \frac{\gamma}{2}\left[\hat{a}\hat{\rho},\hat{a}^\dagger\right] + \frac{\gamma}{2}\left[\hat{a},\hat{\rho}\hat{a}^\dagger\right].
\ee
This equation has been solved by Milburn and Holmes in \cite{milburn1986}.  They used the antinormally ordered quasiprobability distribution known as the ``$Q$-function.''  The $Q$-function allows one to represent $\hat{\rho}$ as a function over the complex plane, and the value of $Q(a)$ is proportional to the probability that $\hat{\rho}$ ``contains'' the coherent state $\ket{a}$.
\be
Q\left(a\right)=\frac{1}{\pi}\langle a|\hat{\rho}|a\rangle.
\ee
Note that $a$ may be complex.  By projecting both sides of the master equation onto $\bra{a}$ from the left and onto $\ket{a}$ from the right, one obtains a differential equation for $Q(a)$.  Milburn and Holmes solved this using a power series and obtained
\begin{eqnarray}
Q_{\mathrm{loss}}(a) = e^{|a|^2} \sum_{q,p=0}^{\infty} \frac{\left(a\beta^\ast\right)^q\left(a^\ast\beta\right)^p}{q!p!} \left(i^{p^2-q^2}\right) \times \nonumber \\
\exp\left[\frac{-\pi\gamma(p+q)}{4\chi}\right] \times \nonumber \\
\exp\left[-|a|^2\left(\frac{\frac{\gamma}{\chi}i^{p-q}e^{-\frac{\gamma}{2\chi}}+2i(p-q)}{\frac{\gamma}{\chi}+2i(p-q)}\right)
\right],
\end{eqnarray}
after $t=\frac{\pi}{2 \chi}$.  $Q_{\mathrm{loss}}$ is a function of $\gamma/\chi$, which is an unitless quantity.

We calculate the fidelity of the state described by $Q_{\mathrm{loss}}$ with the cat in Eq.~(\ref{-icat}) using the relation
\be
F=\pi\int\mathrm{d}^2aP_{\mathrm{cat}}(a)Q_{\mathrm{loss}}(a),
\label{fidelityQP}
\ee
where $P_{\mathrm{cat}}$ is the normally ordered quasiprobability describing the state in Eq.~(\ref{-icat}).  The $P$ function allows us to represent any density matrix as a diagonal sum or integral over coherent states:
\be
\hat{\rho}=\int \mathrm{d}^2\ket{a}\bra{a}P(a).
\ee
We calculate $P_{\mathrm{cat}}(a)$ for Eq.~(\ref{-icat}) using the methods explained in the third chapter of \cite{carmichael1999} and find
\begin{eqnarray}
P_{\mathrm{cat}}(a) = \frac{1}{2}\sum_{m,n=0}^{\infty} \frac{\beta^n\left(\beta^\ast\right)^m}{m!n!} \times \nonumber \\
\left(1+(-1)^{n+m}+ie^{-2|\beta|^2}\left[(-1)^n-(-1)^m\right]\right) \times \nonumber \\
\frac{\partial^n}{\partial a^n}\frac{\partial^m}{\partial (a^\ast)^m}\delta^2(a),
\end{eqnarray}
where $\delta^2(a)$ is the Dirac delta function on the complex plane.

\begin{figure}
\includegraphics[width=8.255cm]{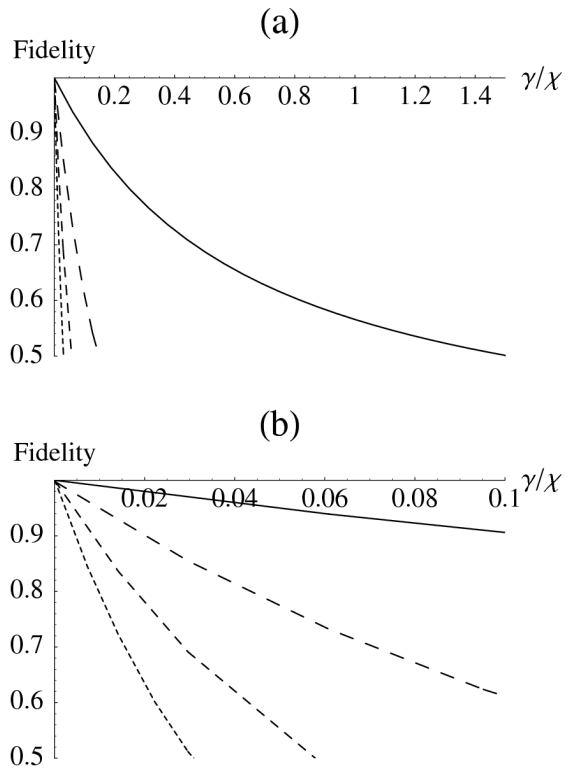}
\caption{(a) Fidelity of the state produced using the Kerr effect in the presence of loss when our goal is to produce a cat of amplitude $\beta=1$ (solid curve), 2 (long dashes), 3 (medium dashes), and 4 (short dashes).  The amplitude of the input coherent state has been optimized to give the maximum fidelity. (b) Same as (a), zoomed in on small $\gamma/\chi$.
\label{fgammachi}}
\end{figure}

We can now evaluate the integral in Eq.~(\ref{fidelityQP}) and find the fidelity with which one can make cat states using a lossy Kerr effect medium.  Plots of the fidelity as a function of $\gamma/\chi$ are in Fig.~\ref{fgammachi}.  Creating a cat of $\beta=1$ with $F>0.5$ requires $\gamma/\chi<1.5$.  Larger cats are much more sensitive to loss.  Making high fidelity cat states directly using the Kerr effect requires some nonlinear optical material/process for which $\gamma/\chi<1$.

Materials exhibiting the Kerr effect are often described using a ``nonlinear index of refraction'' $n_2$, which depends on the light intensity $I$.
\be
n(I) = n_1+n_2I=n_1+n_2\left(\frac{\hbar \omega \hat{n}}{A_{\mathrm{eff}}T}\right),
\ee
where $n_1$ is the usual linear index of refraction, $A_{\mathrm{eff}}$ is the effective cross-sectional area of a pulse of light traveling through the material, and $T$ is the pulse's duration.  To relate $n_2$ to $\chi$ we use
\be
\chi = \frac{\hbar\omega^2n_2}{A_{\mathrm{eff}}T},
\ee
from \cite{kitagawa1986}. To obtain a large $\gamma/\chi$ for a given medium, we should confine the mode's $A_{\mathrm{eff}}$ to be as small as possible.  One way to achieve this would be to use short pulses of $T \sim 1 \ \mathrm{fs}$ traveling through single mode fibers, which typically have $A_{\mathrm{eff}} \sim 7 \mathrm{\mu m}^2$.

Fused silica fibers used for telecommunication have exceptionally low loss of 0.2 dB/km at the wavelength $1550 \ \mathrm{nm}$.  This corresponds to $\gamma = 1.79\times 10^5 \ \mathrm{s}^{-1}$.  Fused silica has a small $n_2=2.6\times10^{-20} \ \mathrm{m}^2/\mathrm{W}$ \cite{agrawal2001}. This corresponds to $\chi=620 \ \mathrm{s}^{-1}$ for these pulses.  Therefore single mode fused silica fibers can give $\gamma/\chi \sim 260$, which is much too large for making cat states.

Chalcogenide (As$_2$S$_3$) glass fibers have larger $n_2=2\times 10^{-18} \ \mathrm{m}^2/\mathrm{W}$.  However, their loss is 100 dB/km at $1.55 \ \mathrm{\mu m}$.  Single mode chalcogenide can have $\gamma/\chi\sim 1.3\times 10^4$.  We have been unable to find any conventional nonlinear optical fibers that can give a smaller $\gamma/\chi$ than that of fused silica \cite{sutherland2003}.

Mecozzi and Tombesi \cite{mecozzi1987} and Yurke and Stoler \cite{yurke1987} published papers in 1987 that proposed making cat-like states using a two mode cross-phase modulating or four wave mixing Kerr effect \cite{mecozzi1987, yurke1987}.  They described the application of the Hamiltonian with the nonlinear interaction term
\begin{equation}
H_{\mathrm{int}}=\chi\left(\hat{a}_1\hat{a}_2^\dagger+\hat{a}_1^\dagger\hat{a_2}\right)^2
\end{equation}
to the input state $\ket{0}_1\ket{\beta}_2$.  After time $\pi/(4\chi)$, the two modes will evolve to
%\begin{equation}
%e^{-i\pi/4}\ket{0}_1\left(\ket{\beta}_2-\ket{-\beta}_2\right)+\left(\ket{-i\beta}_1+\ket{i\beta}_1\right)\ket{0}_2.
%\end{equation}
\begin{equation}
-e^{-i\pi/4}\ket{0}_1\ket{\Psi_{-}(\beta)}_2+\ket{\Psi_{+}(i\beta)}_1\ket{0}_2.
\end{equation}
Here the cat may be found in mode 1 or 2 with probability 0.5, while the other mode contains the vacuum.  Tombesi and Mecozzi presented further analysis in \cite{tombesi1987}, where they discuss both self-phase modulating and cross-phase modulating Kerr effects.

A similar scheme was proposed in 1997 by Vitali, Tombesi, and Grangier \cite{vitali1997a}.  They described using the interaction Hamiltonian
\begin{equation}
H_{\mathrm{int}}=\chi\hat{n}_1\hat{n}_2,
\end{equation}
and the input state $\ket{\alpha}_1\ket{\beta}_2$ to make
\be
\ket{\Psi_{+}(\alpha)}_1\ket{\beta}_2-\ket{\Psi_{-}(\alpha)}_1\ket{-\beta}_2.
\ee
Then one should measure the $x$-quadrature of mode 2.  With probability approximately 0.5 the measurement result will be near $\pm\sqrt{2}\beta$, and mode 1 will be the cat $\ket{\Psi_{\pm}(\alpha)}$.  Vitali, Tombesi, and Grangier also examined this scheme's performance in the presence of loss, concluding that the loss should be very small in order to observe the interference fringes in the $q$-quadrature of the cat state. 

We have not performed a full analysis of photon absorption in the cross-phase modulating Kerr effect.  However, the cat states generated by this method are just as fragile as those generated with the self-phase modulating Kerr effect.  Therefore we expect that the requirement for a low ratio of loss to nonlinear coupling strength in cross-phase modulation is similarly demanding as in the self-phase modulation case.

Two technologies that enhance optical nonlinearities are photonic crystals and electromagnetic induced transparency (EIT).  Photonic crystals may be able to reduce the group velocity of light pulses by factors of 100 to 1000 \cite{soljacic2004,fushman2006}. A 300-fold reduction of the group velocity (compared to $c$) of 1550 nm light was observed in a silicon photonic crystal \cite{vlasov2005}.  Reducing the group velocity would allow one to make an equal size cat with a smaller length of fiber and a proportional reduction of $\gamma$.  However photonic crystal waveguides such as that used in \cite{vlasov2005} have an attenuation of $\sim 3.6 \times 10^{5} \ \mathrm{dB/km}$ \cite{dulkeith2005}, which is much too large to maintain cat state coherence.

EIT is a technique for coupling two (or more) light fields through an atom \cite{fleischhauer2005}. By engineering the interaction between the light fields and the atom, one can obtain many interesting linear and nonlinear optical effects, one of which is cross phase modulation between the light fields \cite{schmidt1996}.  During cross phase modulation, the phase of each light mode advances at a rate proportional to the number of photons in the other mode.  If two coherent states $\ket{\alpha}$ and $\ket{\beta}$ are subjected to cross phase modulation for the correct time, they will evolve to $\ket{\alpha,\beta}+\ket{-\alpha,\beta}+\ket{\alpha,-\beta}-\ket{-\alpha,-\beta}$ \cite{lukin2000}.  This is a state of two cats entangled with one another.  However, cross-phase modulation at the level of only a few photons has not yet been demonstrated \cite{fleischhauer2005}.

\section{Degenerate Optical Parametric Oscillator}
In 1988 Wolinsky and Carmichael presented an analysis of degenerate optical parametric oscillators (DOPO) \cite{wolinsky1988}.  In a DOPO a strong pump field of frequency $2\omega$ is used to pump a nonlinear crystal.  The crystal exhibits a $\chi^{(2)}$ susceptibility and can oscillate at frequency $\omega$ when driven by the pump field.  Through this process pump photons are converted into pairs of signal photons with frequency $\omega$.  To increase the interaction strength the crystal is placed in an optical cavity.  Photons of both frequencies may leak out of the cavity or be absorbed by the crystal.

Wolinsky and Carmichael examined the behavior of a DOPO using the Hamiltonian
\begin{equation}
H=\frac{ig}{2}\left(\left.\hat{a}_s^\dagger\right.^2\hat{a}_p-\hat{a}_s^2\hat{b}_p^\dagger\right)+iE\left(\hat{a}_p^\dagger-\hat{a}_p\right) + H_{\mathrm{loss}},
\end{equation}
where $\hat{a}_s$ is the annihilation operator of the signal mode, and $\hat{a}_p$ is the annihilation operator of the pump mode.  $H_{\mathrm{loss}}$ describes the linear losses from the cavity mirrors and in the crystal.  When the pump is treated as a classical field, and loss is neglected, this Hamiltonian produces squeezed light.  Wolinsky and Carmichael analyzed this system, including loss, in the strong coupling regime with $g\sim 1$, which is much larger than that achieved in current squeezing experiments.  Their analysis was based on the positive-$P$ representation, an approximation of the usual normally ordered $P$ representation.  Their analysis suggested that in the regime where the ratio of the pump strength $\lambda$ to $g^2$ is very small, one may find the cat state $\left|\sqrt{\lambda}/g\right\rangle_s+\left|-\sqrt{\lambda}/g\right\rangle_s$ in this system.

However, Reid and Yurke \cite{reid1992} showed that for any finite number of photons in the signal mode, the steady state of the degenerate parametric oscillator in the strong coupling regime is instead an incoherent mixture of $\left|\sqrt{\lambda}/g\right\rangle_s$ and $\left|-\sqrt{\lambda}/g\right\rangle_s$.  The system's Wigner function is always positive, and no interference fringes would be visible in homodyne detection.  Further analysis by Krippner, Munro, and Reid \cite{krippner1994} investigated the short-time evolution of this system starting with the signal mode in the vacuum state.  Their calculations indicate that the signal mode may exhibit interference fringes indicative of a cat-like state, but these fringes only survive for times much shorter than the signal mode's cavity decay time.  Unfortunately, the $\chi^{(2)}$ coupling strength required to produce cats using this method is vastly larger than the strength of currently available materials.  Also, we do not have a good method to confirm the cat's presence, because it is trapped in the optical cavity and will decohere when it exits the cavity mirrors.

\section{Back-Action Evasion Measurement}

\begin{figure}
\includegraphics[width=8.255cm]{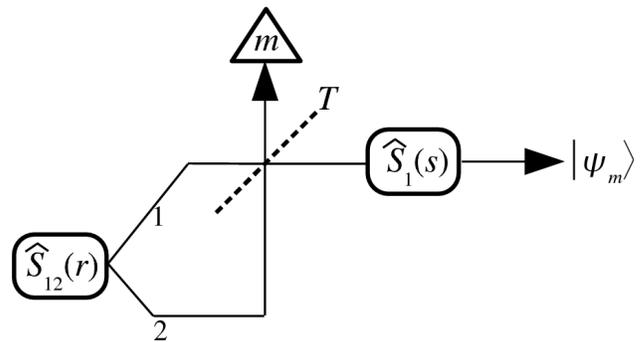}
\caption{Diagram of the scheme to make cats using the back-action evasion measurement. $\hat{S}_{12}(r)$ is a nondegenerate down conversion crystal, which creates squeezed light in modes 1 and 2 at frequency $\omega$.  The dashed line is a beam splitter with transmissivity $T$.  The triangle is a photon counter, and $\hat{S}_1(s)$ is a degenerate down conversion crystal that squeezes mode 1.  Not shown are a light beam or beams used to pump the down conversion processes and mirrors to redirect the light. \label{song_scheme}}
\end{figure}

Because of the great difficulty in creating cats directly using the Kerr effect, researchers have proposed numerous methods for making cats based on squeezing followed by measurement and postselection.  The first of these was described by Song, Caves, and Yurke in \cite{song1990}.  Their scheme is in Fig.~\ref{song_scheme}.  It is based on the back-action evasion measurement described in \cite{yurke1985}, except that the input state is the vacuum, and we measure the number of photons rather than one of the quadratures.

We begin by making an entangled, squeezed state of modes 1 and 2 through non-degenerate down-conversion.  This prepares the state
\be
\hat{S}_{12}(r)\ket{0_1,0_2}=e^{r\left(\hat{a}_1\hat{a}_2-\hat{a}_1^\dagger \hat{a}_2^\dagger\right)}\ket{0_1,0_2}.
\ee
The wavefunction describing the $x$-quadratures for the two modes is
\begin{eqnarray}
&\psi_{\mathrm{sq}}(x_1,x_2)=\langle x_1,x_2|\hat{S}_{12}(r)\ket{0_1,0_2} = \nonumber \\
& \frac{1}{\sqrt{\pi}}\exp\left[-\frac{1}{2}e^{-2r}\left(\frac{x_1}{\sqrt{2}}-\frac{x_2}{\sqrt{2}}\right)^2 -\frac{1}{2}e^{2r}\left(\frac{x_1}{\sqrt{2}}+\frac{x_2}{\sqrt{2}}\right)^2\right]. \nonumber \\
&
\end{eqnarray}
Modes 1 and 2 then meet in a beam splitter with transmissivity $T$, which performs the transformation $x_1 \rightarrow \sqrt{T}x_1+\sqrt{1-T}x_2$ and $x_2 \rightarrow \sqrt{1-T}x_1-\sqrt{T}x_2$ on $\psi_\mathrm{sq}(x_1,x_2)$.  We next count the number of photons in mode 2, obtaining the result $m$.  This projects the state onto the $m$ photon eigenstate of mode 2, which is 
\be
\phi_m(x_2)=e^{-x^2/2}\frac{H_m(x_2)}{\sqrt{2^m m!\sqrt{\pi}}},
\label{photon_number_eigenstate}
\ee
where $H_m(x)$ is the Hermite polynomial with index $m$.
Last, mode 1 is squeezed in the degenerate down-conversion 
\be
\hat{S}_1(s)=\exp\left[\frac{s}{2}\left(\hat{a}_1^2-\left(\hat{a}^\dagger_1\right)^2\right)\right]. 
\ee
The final state is
\begin{eqnarray}
\psi_m(x_1)  &=&  \frac{e^{s/2}}{\sqrt{\mathrm{P}(m)}}\int_{-\infty}^\infty \mathrm{d}x_2\phi_m(x_2) \nonumber \\
& \times & \psi_\mathrm{sq}(\sqrt{T}x_1e^s+\sqrt{1-T}x_2,\sqrt{1-T}x_1e^s-\sqrt{T}x_2),\nonumber \\
& &
\end{eqnarray}
where $\mathrm{P}(m)$ is the probability to measure $m$ photons.  We will use $\psi_m$ and $\ket{\psi_m}$ to denote the output of several schemes to make cats by post selecting on counting $m$ photons, but in each case $\psi_m$ will represent a different state.

We have performed the first calculations of the fidelity of the state produced from the back-action evasion scheme with the perfect cat state wave function $\Psi_\pm^{(\alpha)}(x)$ from Eq.~(\ref{cat_x_wavefunction})
\be
F=\left|\int_{-\infty}^\infty \mathrm{d}x_1\Psi_\pm^{(\alpha)}(x)\psi_m(x_1)\right|^2.
\ee
To maximize the fidelity with which we can make a particular cat, we optimize over the two squeezing parameters $r$ and $s$, the beam splitter transmissivity $T$, and the number of photons detected $m$.  [In the original scheme of Song, Caves, and Yurke, $T=\cos^2(1/2\arcsin(\tanh r))$, which is the value of $T$ required for a back-action evasion measurement, but here we treat $T$ as a free parameter.]  For a particular $m$, wegenerally find a significant region of the $(r,s,T)$ parameter space that gives the maximum fidelity.  

Table \ref{table_song_fidelity_2} shows the fidelities with which we can make an even cat state with $\alpha=2$, along with the parameters that give the maximum fidelity.  Particularly we choose to show the combination $(r, s, T)$ that gives the highest $\mathrm{P}(m)$ with which one can make the highest fidelity cat states.  By increasing $T$ and reducing $r$ and $|s|$, one can obtain cats with the same fidelity but lower probability.  The table shows fidelities only for even $m$, because for odd $m$ the fidelity is always 0.

\begin{table}[h]
{\bf \caption{\label{table_song_fidelity_2} Fidelities for Making Cats with Back-Action Evasion}}
\begin{center}
\begin{tabular}{llllll}
\hline
	$m$\hspace{5mm} & $F$\hspace{10mm} & $\mathrm{P}(m)$\hspace{5mm}  & $r$\hspace{10mm}  & $s$\hspace{10mm} & $T$ \\
\hline
	2 & 0.9709 & 0.110 & 1.14 & -1.35 & 0.808\\
	4 & 0.9978 & 0.056 & 1.44 & -1.48 & 0.710\\
	6 & 0.9995 & 0.038 & 1.61 & -1.63 & 0.652\\
	8 & 0.9998 & 0.029 & 1.76 & -1.77 & 0.616\\
\hline
\end{tabular}
\end{center}
\begin{flushleft}
\footnotesize Fidelities with which one can make even cat states with $\alpha=2$ using the back-action evasion scheme.  We optimize $r$, $s$, and $T$ first, to give highest fidelity, and second, to give highest probability. \normalsize
\end{flushleft}
\end{table}

Let us discuss the case of $m=2$ in more detail.  The values shown in the table show the highest probability with which one can obtain the highest fidelity cat states.  We can reduce the demand for high squeezing by accepting a lower success rate.  For example, using $(r=0.00235, s=-0.588, T=0.999999)$, we obtain cats with the same $F=0.9709$ with probability $\mathrm{P}(2)=4\times10^{-11}$.  A reasonable compromise could be $(r=0.074,s=-0.593, T=0.999)$ at $\mathrm{P}(2)=4\times10^{-5}$.  This gives flexibility to reduce the required level of squeezing while suffering a lower success probability.

\begin{figure}
\includegraphics[width=8.255cm]{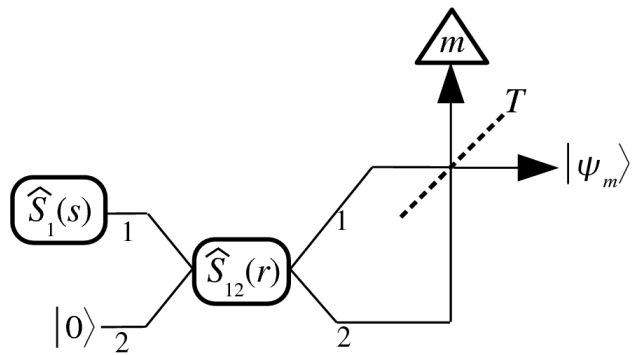}
\caption{Diagram of the improved scheme to make cats using the back-action evasion measurement. It is similar to that in Fig.~\ref{song_scheme} except that $\hat{S}_1(r)$ has been moved to the beginning of the network.  Not shown are a light beam or beams used to pump the down conversion processes and mirrors to redirect the light.
\label{yurke_scheme}}
\end{figure}

The original back-action evasion scheme of Song, Caves, and Yurke was improved by Yurke, Schleich, and Walls in \cite{yurke1990}.  They recommended squeezing mode 1 before the two mode squeezing $\hat{S}_{12}(r)$.  This removes the need to squeeze mode 1 after the beam splitter used for the back-action evasion measurement, and it reduces the need for strong squeezing.  Figure \ref{yurke_scheme} shows the new procedure.

We analyzed this improved scheme using methods similar to those just described for the original back-action evasion scheme.  We again show a table of fidelities for making an even cat state with $\alpha=2$ in Table \ref{table_yurke_fidelity}.  Moving the squeezing of mode 1 to the beginning of the protocol allows us to produce cats with equal fidelity with much lower levels of squeezing.  For the $m=4$, 6, and 8 we can remove the beam splitter and count the photons in mode 2 just after the entangling operation of $\hat{S}_{12}(r)$.  One can make cats with fidelities equal to those shown in the table requiring less squeezing, by accepting a lower $P(m)$.  For example one may use the combination $m=2$, $r=-0.0009$, $s=-0.589$, and $T=0.547$, to make an $\alpha=2$ even cat state with probability $P(m)=1.01\times10^{-6}$ and $F=0.9709$.

\begin{table}[h]
{\bf \caption{\label{table_yurke_fidelity} Fidelities for Making Cats with the Improved Back-Action Scheme}}
\begin{center}
\begin{tabular}{llllll}
\hline
	$m$\hspace{5mm} & $F$\hspace{10mm} & $\mathrm{P}(m)$\hspace{5mm} & $r$\hspace{10mm} & $s$\hspace{10mm} & $T$ \\
\hline
	2 & 0.9709 & 0.110 & -0.263 & -1.36 & 0.972\\
	4 & 0.9978 & 0.056 & -0.271 & -1.41 & 1\\
	6 & 0.9995 & 0.0017 & -0.162 & -0.62 & 1\\
	8 & 0.9998 & 0.000016 & -0.116 & -0.45 & 1\\
\hline
\end{tabular}
\end{center}
\begin{flushleft}
\footnotesize
Fidelities with which one can make even cat states with $\alpha=2$ using the improved back-action evasion scheme shown in Fig.~\ref{yurke_scheme}.  We optimize $r$, $s$, and $T$ first, to give highest fidelity, and second, to give highest probability.
\normalsize
\end{flushleft}
\end{table}

\begin{figure}
\includegraphics[width=8.255cm]{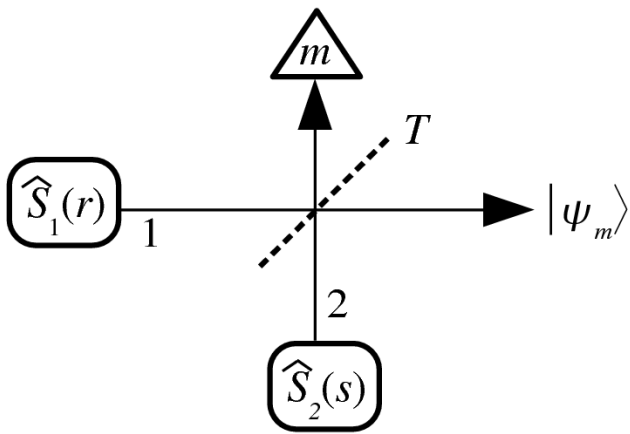}
\caption{Diagram of our simplified ``back-action evasion'' scheme to make cats. Modes 1 and 2 begin in the vacuum state, and are transformed to squeezed states by $\hat{S}_1(r)$ and $\hat{S}_2(s)$.  The two squeezed states meet at the beam splitter with transmissivity $T$.  We then count the number of photons in mode 2, obtaining result $m$.  Mode 1 then contains $\ket{\psi_m}$, which should be similar to a cat state.  Not shown are a light beam or beams used to pump the down conversion processes and mirrors to redirect the light.
\label{glancy_scheme}}
\end{figure}

We now introduce a new simplification of the ``back-action evasion'' scheme, which is now quite divorced from the original back-action evasion ideas of \cite{yurke1985}.  Because the beam splitter used to interfere modes 1 and 2 before the photon counting measurement shown in Fig.~\ref{yurke_scheme} is unnecessary for $m \geq 4$, we remove it altogether.  One can always turn a two mode squeezing operation into two single mode squeezers followed by a beam splitter \cite{braunstein2005}: $\hat{S}_{12}(r)=\hat{B}_{12}(1/2)\hat{S}_2(-r)\hat{S}_1(r)$.  Therefore the sequence $\hat{S}_{12}(r)\hat{S}_1(s)=\hat{B}_{12}(1/2)\hat{S}_2(-r)\hat{S}_1(s+r)$, which is just independent squeezing of modes 1 and 2 followed by a beam splitter.  This leads us to consider a scheme that begins with squeezed states in both modes 1 and 2, which then interfere at a beam splitter with transmissivity $T$.  We count the number of photons in mode 2, obtaining result $m$ and preparing the output of mode 1 in the state $\ket{\psi_m}$, which should be similar to a cat.  Figure \ref {glancy_scheme} contains a diagram of this scheme.

After having been squeezed, modes 1 and 2 are in the state
\begin{equation}
\psi_{\mathrm{sq}}(x_1)\psi_{\mathrm{sq}}(x_2)=\frac{e^{\frac{r}{2}+\frac{s}{2}}}{\sqrt{\pi}}\exp\left[-\frac{(x_1e^r)^2}{2}-\frac{(x_2e^s)^2}{2}\right].
\end{equation}
The beam splitter causes the the rotation of $x_1$ and $x_2$ described above.  After measuring $m$ photons in mode 2, the output state is
\begin{eqnarray}
\psi_m(x_1) & = & \int_{-\infty}^\infty \mathrm{d}x_2 \phi_m(x_2) \psi_{\mathrm{sq}}\left(\sqrt{T}x_1+\sqrt{1-T}x_2\right) \nonumber \\
& & \times \psi_{\mathrm{sq}}\left(\sqrt{1-T}x_1-\sqrt{T}x_2\right),
\end{eqnarray}
where $\phi_m(x_2)$ is the photon number eigenstate wave function from Eq. (\ref{photon_number_eigenstate}).  We show a table with the fidelity with which this state approximates an even cat state of $\alpha=2$ in Table \ref{glancy_table}.  Again in this scheme we can trade off squeezing strength for lower $P(m)$ while keeping the same fidelity.  We could make an even cat state of $\alpha=2$ whose fidelity is 0.9709 using $m=2$, $r=-0.590$, $s=0.0010$, and $T=0.9975$ with $P(2)=1.3\times10^{-6}$.  Here the squeezing of mode 2 is so small, that we are tempted to remove it altogether.  We discuss such a scheme in detail in the next section.

\begin{table}[h]
{\bf \caption{\label{glancy_table} Fidelities for Making Cats with Simplified Back-Action Scheme }}
\begin{center}
\begin{tabular}{llllll}
\hline
	$m$\hspace{5mm} & $F$\hspace{10mm} & $\mathrm{P}(m)$\hspace{5mm} & $r$\hspace{10mm} & $s$\hspace{10mm} & $T$ \\
\hline
	2 & 0.9709 & 0.110 & 0.263 & -1.62 & 0.665\\
	4 & 0.9978 & 0.056 & 0.274 & -1.76 & 0.497\\
	6 & 0.9995 & 0.038 & 0.221 & -1.85 & 0.398\\
	8 & 0.9998 & 0.029 & 0.182 & -1.93 & 0.332\\
\hline
\end{tabular}
\end{center}
\begin{flushleft}
\footnotesize
Fidelities with which one can make even cat states with $\alpha=2$ using our simplified ``back-action evasion'' scheme shown in Fig.~\ref{glancy_scheme}.  We optimize $r$, $s$, and $T$ first, to give highest fidelity, and second, to give highest probability.
\normalsize
\end{flushleft}
\end{table}

Two difficulties for implementing these ``back-action evasion'' schemes are their need for a high efficiency photon counter and strong squeezing.  If the photon counter misses one photon, the scheme makes a state orthogonal to the desired cat, so the fidelity would be significantly degraded.  Experimentalists often report results of squeezing experiments using the number of decibels of the variance of the quadrature noise above or below the variance of the vacuum's noise.  The number of decibels is related to our squeezing parameter $s$ by
\be
\mathrm{\# \ of \ dB}= - 10 \log_{10}e^{-2s}\approx8.69 s
\ee
for the $x-$quadrature, and the negative of this for the $p-$quadrature.  A squeezing parameter of $s=1$ corresponds to $\sim -8.7$ dB of quadrature noise power squeezing below the shot noise limit.  The most impressive squeezing experiments for continuous wave light that the authors are aware of have achieved $-9$ dB \cite{takeno2007} and $-10$ dB \cite{vahlbruch2007} of squeezing.  In a typical pulsed squeezing experiment one may obtain $-3$ dB ($s=0.35$) of squeezing \cite{wenger2004a}.  Fortunately these schemes allow a trade-off in which one may use weaker squeezing to make equally high fidelity cats with a decreased success probability.

Here we have made no effort to analyze any impurity or loss in the squeezing, which is likely to have significant effects.
For discussion of loss we refer the reader to a pair of papers by Tombesi and Vitali \cite{tombesi1996} and Vitali and Tombesi \cite{vitali1997}.  In addition to loss, these papers describe a method for detecting the cat state by exploiting the back-action evasion interaction.  They imagine producing a cat in a cavity that has much lower loss for the signal (cat-containing) mode than for the meter mode.  The meter mode exits the cavity, and we count its photons.  This prepares the signal mode in a cat state.  We could then use the back-action evading interaction between the signal and meter modes to measure the signal's state by performing homodyne detection of the meter's quadratures.  Using this method requires that the back-action evasion conditions be met, so one could not freely optimize $T$ for making high fidelity cats.

It is also well known that highly squeezed states are very sensitive to phase noise \cite{takeno2007}.  It would be very challenging to incorporate the high efficiency photon counter, two strong down conversion processes, and stable phase control in a single experiment.

\section{Photon Subtraction}
\begin{figure}
\includegraphics[width=8cm]{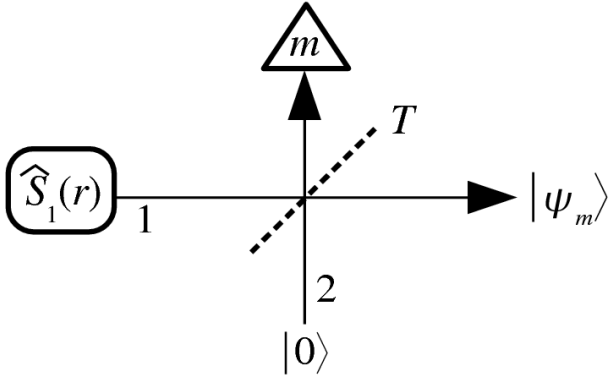}
\caption{Diagram for the generation of cat states by means of a conditional photon number measurement on a beam splitter. The down conversion $\hat{S}_1(r)$ creates the single mode squeezed state in mode 1.  It is input into one port of a variable transmissivity $T$ beam splitter with mode 2 containing a vacuum state. A definite measurement of $m$ photons on one output port of the beam splitter prepares the state $|\psi_m\rangle$, which is a good approximation to a cat state.
\label{dakna_scheme}}
\end{figure}

A simplification of the back-action evasion scheme was proposed by Dakna and co-authors \cite{dakna1997}.  It too uses squeezing followed by photon counting but accomplishes similar goals with only one stage of squeezing.  Their scheme is depicted in Fig.~\ref{dakna_scheme} and works as follows. First it requires first the preparation of a squeezed state of light. This can be accomplished by sending a laser beam at twice the frequency of the desired squeezed state through a down-converting crystal. This prepares mode 1 in the state
\begin{eqnarray}
|\psi_{sq}\rangle & = & e^{r\left(\hat{a_1}^2-\left(\hat{a_1}^{\dagger}\right)^2\right)/2}|0\rangle
\nonumber\\
 & = & \left(1-\lambda^2 \right)^{\frac{1}{4}} \sum_{n=0}^{\infty} \frac{
 \sqrt{(2n)!}}{n!} \left(\frac{\lambda}{2}\right)^n | 2 n\rangle,
\end{eqnarray}
where $\lambda = -\tanh r$. 
$|\psi_{sq}\rangle_1$ has only even numbers of photons, so in this respect it is already like the even cat state.

To make a cat, we combine the squeezed state and a vacuum state on a beam splitter $\hat{B}_{12}(T)$ with variable transitivity $T$. On the mode 2 output port from the beam splitter we count the number of photons and obtain the result $m$. The conditional state of output mode 1 is then
\begin{eqnarray}
\label{daknacat}
|\psi_m\rangle&=& \frac{1}{\sqrt{\cal N}_m} \sum_{n=0}^{\infty} c_{n,m} \left(\frac{\lambda T}{2}\right)^\frac{n+m}{2} | n\rangle,
\end{eqnarray}
where
\begin{eqnarray}
c_{n,m}=
\frac{
(n+m)!\left(1+\left(-1\right)^{n+m}\right)
}{
\left(\sqrt{n!}\Gamma \left(\frac{n+m}{2}+1\right)\right)
}
\end{eqnarray}
and
${\cal N}_m= \sum_n c_{n,m}^2 \left( \frac{\lambda T}{2}\right)^{n+m}$.  If an odd number of photons $m$ is registered in the 
detector, then $c_{n,m}=0$ for all even $n$'s, and if $m$ is even, then $c_{n,m}=0$ for all odd $n$'s. This condition is required by photon 
conservation; because $|\psi_{sq}\rangle_1$ has only even numbers of photons, the total number of photons emerging from the output ports of the 
beam splitter must also be even. The mean photon number of $|\psi_m\rangle$ is
\begin{eqnarray}
\langle \bar n \rangle=\frac{1}{{\cal N}_m} \sum_{n=0}^{\infty} n c_{n,m}^2
\left( \frac{\lambda T}{2}\right)^{n+m}.
\end{eqnarray}

Eq.~(\ref{daknacat}) can be broken into two cases: the state resulting from an even $m$ result (which should be similar to an even cat state) 
and the state from an odd $m$ (which should be similar to an odd cat). For $m$ even, Eq.~(\ref{daknacat}) has only even photon numbers and 
can be written
in the simplified form
\begin{eqnarray}
\label{evendaknacat}
|\psi_m\rangle&=& \frac{1}{\sqrt{\cal N}_m} \sum_{n=0}^{\infty} \frac{(2 n+m)!
\left(\frac{\lambda T}{2}\right)^{n+\frac{m}{2}} }
{ \left(n+\frac{m}{2}\right)! \sqrt{(2n)!}} | 2 n \rangle.
\end{eqnarray}
For $\lambda T$ small, this expression can be further approximated as
\begin{eqnarray}
|\psi_m\rangle&\approx& | 0 \rangle  + \lambda T
\frac{1+m}{\sqrt{2}}| 2 \rangle+\ldots.
\end{eqnarray}
As $m$ increases, so does the population in the $| 2 \rangle$ (and higher) states compared with the $m=0$ situation. Thus for small $\lambda T$
the mean photon number increases as $m$ increases.

\begin{figure}
\includegraphics[width=8.255cm]{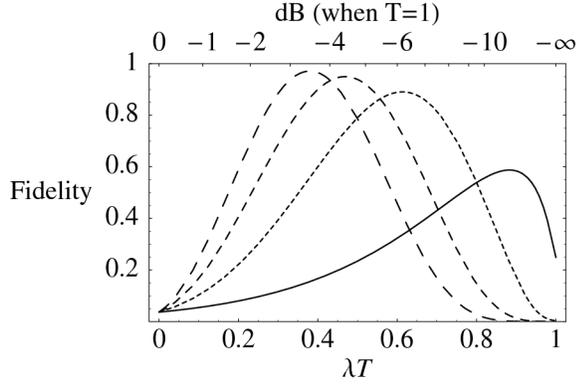}
\caption{Plot of the fidelity of the state $|\psi_m\rangle$ with $|\Psi_+(2)\rangle$ versus $\lambda T$ for m=0 (solid curve), m=2 (small dashes), m=4 (medium dashes), m=6 (large dashes).  The upper horizontal axis shows the number of decibels of squeezing required when $T=1$ is used.
\label{daknafidelitylambdat}}
\end{figure}

To find how closely the states $|\Psi_m\rangle$ match true even cats, we calculate the 
fidelity $F=|\langle\Psi_+(\alpha) |\psi_m\rangle|^2$ between the two states. Plots of the fidelity of $|\Psi_+(2)\rangle$ versus $\lambda T$ for various
$m$'s appear in Fig.~\ref{daknafidelitylambdat}.  One can clearly see that the fidelity is improved and the amount of required squeezing is lower when more photons are detected. 

\begin{figure}
\includegraphics[width=8.255cm]{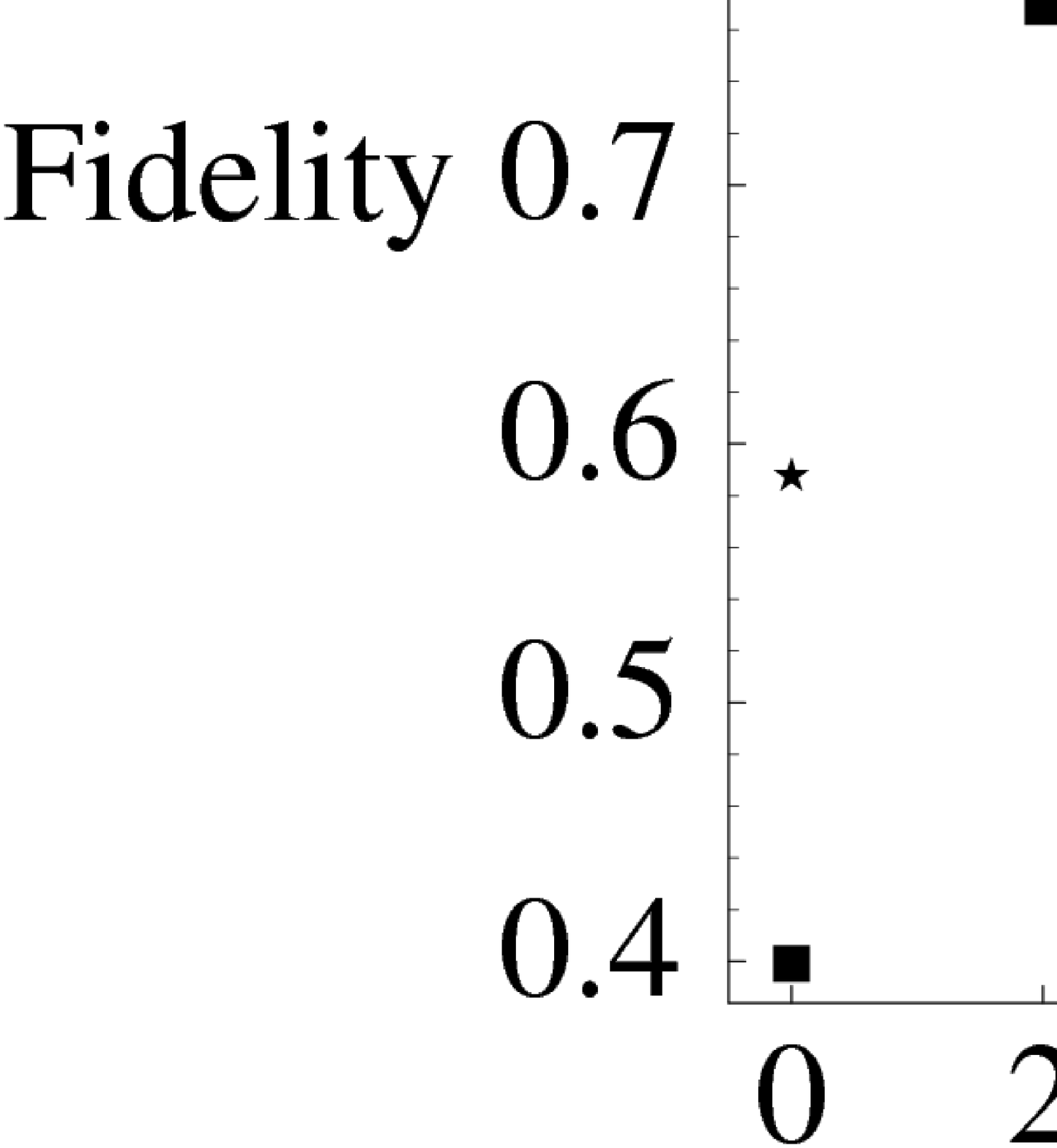}
\caption{Plot of the fidelity of the state $|\Psi_m\rangle$ with $|\Psi_+(\alpha)\rangle$ for $\alpha=1$ (diamonds), $2$ (stars), and $3$ (squares).  Notice that only even $m$ are represented, because odd $m$ events give a fidelity of zero.  For each point we have numerically 
optimized $\lambda T$ to give the maximum fidelity.
\label{daknafidelitym}}
\end{figure}

\begin{figure}
\includegraphics[width=8.255cm]{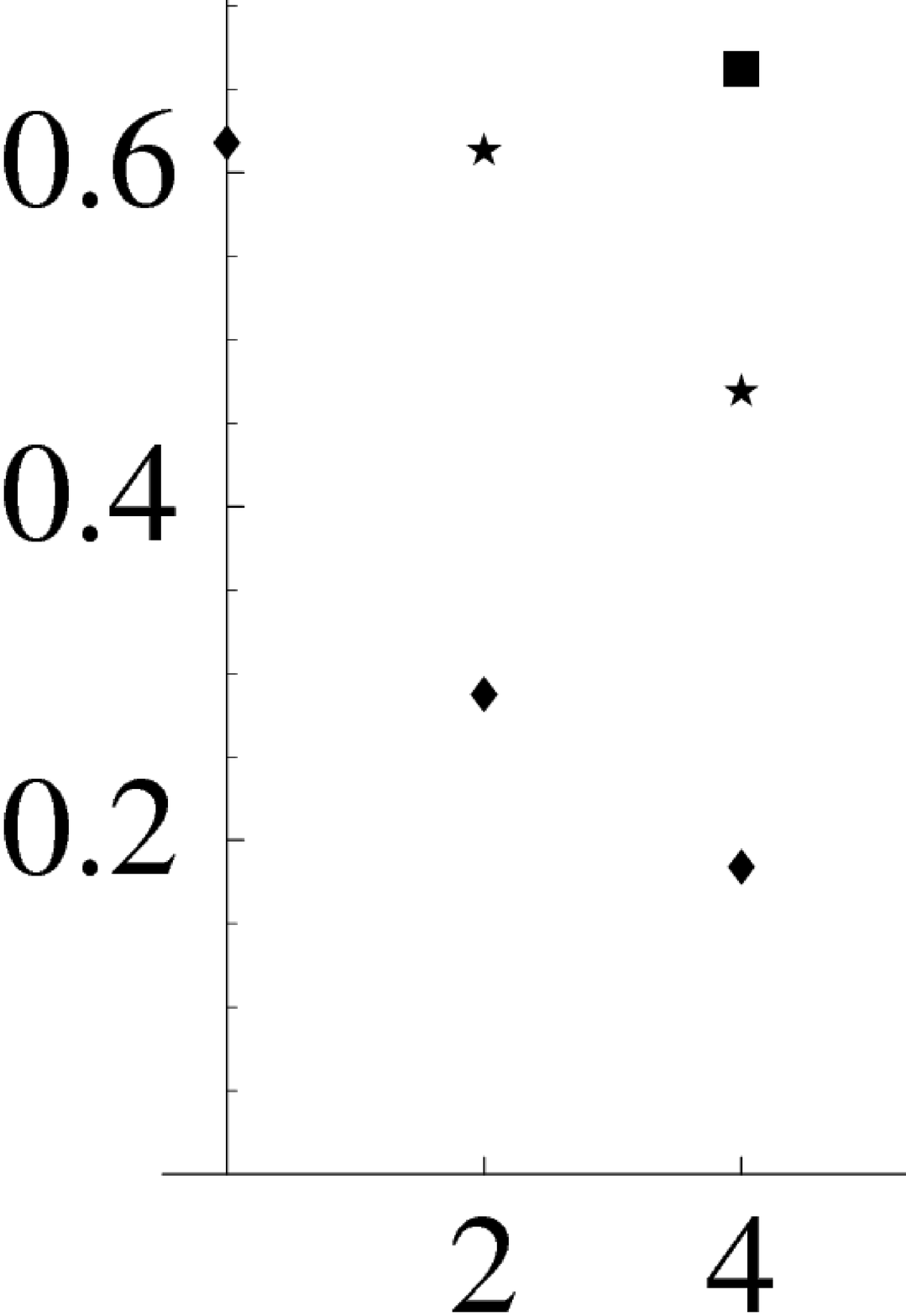}
\caption{Plot of the product $\lambda T$ that maximizes the fidelity shown in Fig.~\ref{daknafidelitym}.
\label{daknalambdatm}}
\end{figure}

We now examine how the maximum fidelity (maximized over the values of $\lambda T$) increases with $m$.  In Fig.~\ref{daknafidelitym} 
we plot the maximum fidelity for each $m$ measurement when attempting to make even cats with $\alpha = 1, 2, 
\text{ and } 3$.  Fig.~\ref{daknalambdatm} shows the $\lambda T$ that optimizes the fidelity for each $m$ measurement.

Making a high fidelity cat state of a given $\alpha$ is best achieved by conditioning successful cat production on detecting 
larger numbers of photons and using smaller $\lambda T$.  Unfortunately, large $m$ detection events occur less frequently as the amount 
of squeezing is decreased.  The probability of detecting $m$ photons is

\begin{eqnarray}
\label{probdaknacat}
{\rm P}(m) & = & \sqrt{\frac{1-\lambda^2}{1-(\lambda T)^2}}\left[\frac{\lambda^2 T(1-T)}{1-(\lambda T)^2} \right]^m
\nonumber \\
 & & \times \sum_{l=0}^{\rm Int[m/2]} \frac{m!}{(m-2 l)!l!^2(2 \lambda T)^{2l}}.
\end{eqnarray}

\begin{figure}
\includegraphics[width=8.255cm]{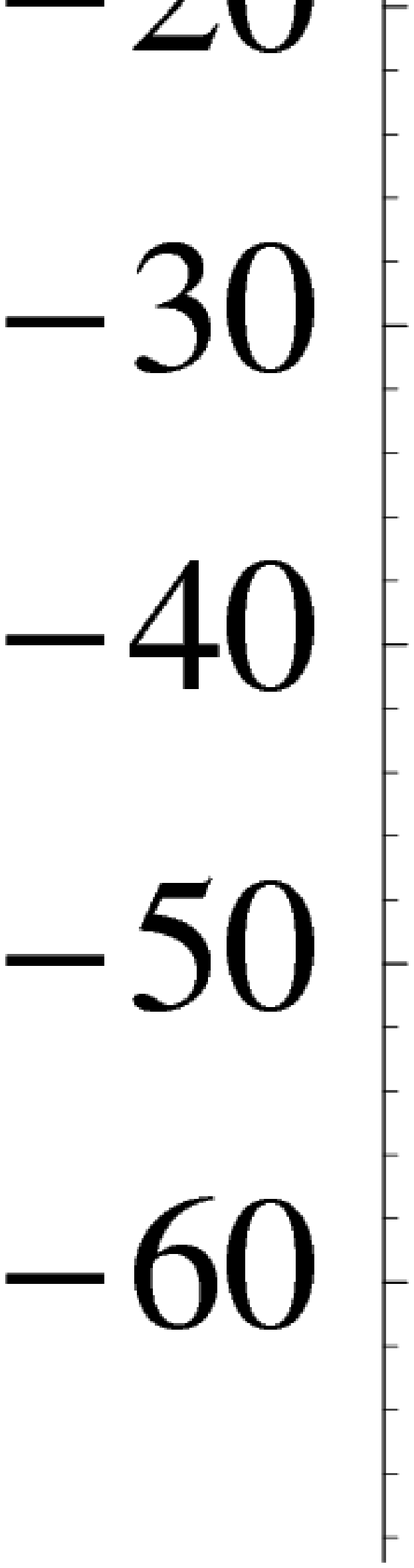}
\caption{Plots of the probability to detect $m$ photons after sending a squeezed vacuum state with squeezing parameter $\lambda$ through a beam splitter with transmissivity $T$. Each plot shows the probability obtained using the product $\lambda T$ that maximizes the fidelity shown in Fig.~\ref{daknalambdatm}.  Plot (a) shows the case in which $\lambda=0.99$ and the probability is high. Plot (b) shows the case in which $\lambda=T$. Plot (c) shows the case in which squeezing is decreased and $T=0.99$.
\label{dakna_probability_m}}
\end{figure}

Although the state $|\psi_m\rangle$ and its fidelity with an even cat depend only on the product $\lambda T$, its probability 
depends on $\lambda$ and $T$ separately.  ${\rm P}(m)$ can be increased by increasing $\lambda$ while decreasing $T$ (while keeping 
$\lambda T$ fixed), but producing highly squeezed light is difficult. Fig.~\ref{dakna_probability_m} plots the probability to detect $m$ 
photons for various $\lambda $ and $T$ combinations.

\begin{figure}
\includegraphics[width=8.255cm]{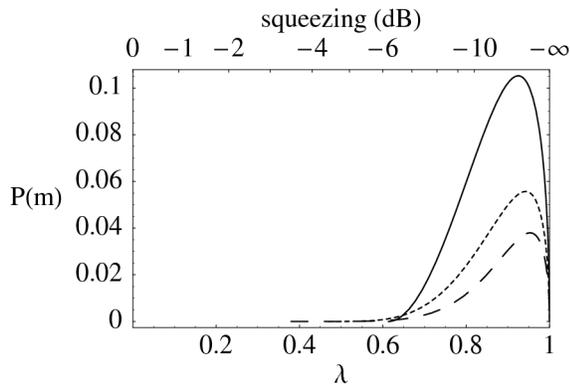}
\caption{Probability to detect m=2 (solid curve), 4 (small dashes), and 6 (long dashes) photons as a function 
of squeezing (shown with both $\lambda$ and in decibels).  In each case $T$ is adjusted so that $\lambda T$ gives the best fidelity to produce a cat state with $\alpha=2$.
\label{dakna_probability_lambda}}
\end{figure}

Let us examine a few illustrative cases.  Suppose we desire to create an even cat state with $\alpha=2$.  The squeezed state closest to the 
$\alpha=2$ cat has $\lambda=0.883$ and a fidelity of only $0.588$.  If we detect $m=2$ photons from the state prepared using 
$\lambda T=0.613$, we will create a cat with a fidelity of $0.891$.  Detecting $m=4$ using $\lambda T=0.469$ produces a fidelity of $0.950$.  
Detecting $m=6$ using $\lambda T=0.380$ produces a fidelity of $0.971$.  Notice that higher fidelity is achieved by detecting larger numbers 
of photons and using smaller $\lambda T$.  We would like to select $\lambda$ and $T$ to give the highest success probability.  In Figure 
\ref{dakna_probability_lambda} we plot ${\rm P}(m)$ versus $\lambda$ for detecting $m=2$, $4$, and $6$ photons.  The beam splitter's 
transmissivity is adjusted to keep $\lambda T$ equal to the value giving the best fidelity for each photon detection case.

A weakness of this method for creating cat states is its need for high efficiency photon counters.  If the detector misses one photon then this scheme will create a state that is orthogonal to the desired state.  Suppose, for example, we wish to make an even cat after 
detecting $m=2$, but three photons actually arrive at the detector.  If the detector registers only two photons, then we have produced an odd cat, but 
we falsely believe we have an even cat.  Other researchers have analyzed photon detector inefficiency in photon subtraction for the case of subtracting only one photon or using photon detectors that cannot discriminate 1, 2, ... photons \cite{ourjoumtsev2006,kim2005,olivares2005}.  Here we will consider true photon counting.  Let the 
detector's efficiency be represented by $\eta$.  The probability to register $m$ photons (meaning that the detector reports seeing $m$ 
photons, when more than $m$ may actually arrive at the detector) is given by Eq.~(\ref{inefficiency_transformation}).  The state produced when 
the detector registers $m$ photons is now given by the density operator

\begin{equation}
\rho_{\eta}(m)=\frac{1}{{\rm P}_{\eta}(m)} \sum_{n=m}^{\infty} {\rm P}(n) \binom{n}{m} \eta^m (1-\eta)^{n-m} |\Psi_n\rangle \langle \Psi_n|.
\end{equation}

\begin{figure}
\includegraphics[width=8.225cm]{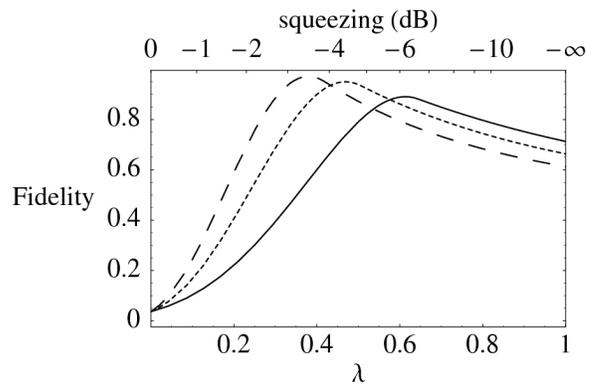}
\caption{Fidelity to make a cat state with $\alpha=2$ as a function of $\lambda$ when using detectors with 
efficiency $\eta=0.9$.  The solid curve shows the $m=2$ case, the small dashed curve shows the $m=4$ case, and the large dashed curve 
shows the $m=6$ case.  For each combination of $\lambda$ and $m$, the beam splitter transitivity has been adjusted to give the maximum 
fidelity.  These maximum fidelities are equal to the maximum fidelities achievable perfect detectors are used.
\label{dakna_eta_fidelity_lambda}}
\end{figure}

\begin{figure}
\includegraphics[width=8.255cm]{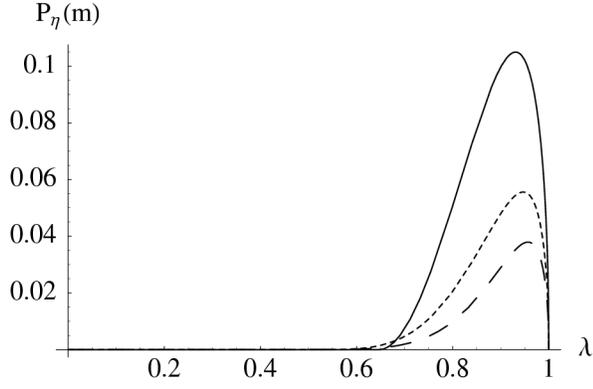}
\caption{Probability to detect $m=2$ (solid curve), $m=4$ (small dashes), and $m=6$ (long dashes) as a function of $\lambda$ when detectors with efficiency $\eta=0.9$ are used.  For each combination of $\lambda$ and $m$, the beam splitter transitivity has been adjusted to give the maximum fidelity, as shown in Fig.~\ref{dakna_eta_fidelity_lambda}.
\label{dakna_eta_probability_lambda}}
\end{figure}

\begin{figure}
\includegraphics[width=8.255cm]{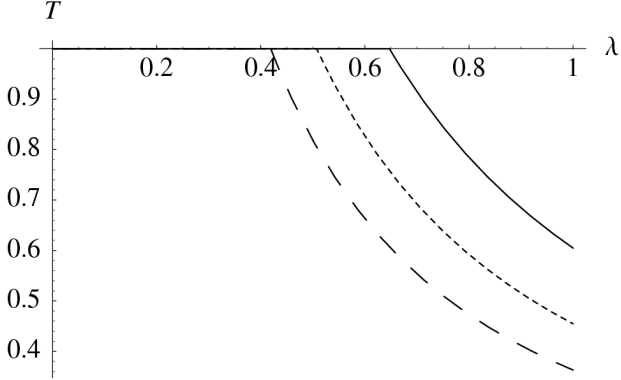}
\caption{Beam splitter transitivity $T$ that gives the largest maximum fidelity in Fig.~\ref{dakna_eta_fidelity_lambda} as a function of $\lambda$.  The solid curve shows the $m=2$ case, the small dashed curve shows the $m=4$ case, and the large dashed curve shows the $m=6$ case.
\label{dakna_eta_T_lambda}}
\end{figure}

The fidelity of $\rho_{\eta}(m)$ depends on the probability distribution given by ${\rm P}(n)$.  In fact given that $m$ photons are 
registered, the fidelity will be greatest when ${\rm P}(m+1) \ll {\rm P}(m)$.  When this condition holds, we can be confident that the 
detector's registering $m$ photons is not caused by the arrival of more than $m$ photons and the detector's inefficiency.  This condition can 
be achieved by using less squeezing and a beam splitter with larger $T$.  Of course, this will result in a decrease in the probability to 
register $m$ photons.  By choosing the best combination of $\lambda$ and $T$ we can obtain a fidelity that is equal to that achievable with perfect detectors.  Unfortunately, with the optimal combination of $\lambda$ and $T$, $\mathrm{P}(m)$ is very small.  Let us continue our examination of the $\alpha=2$ case, but now the total efficiency of the photon counters and other optical elements in the photon counting mode is $\eta=0.9$.  We would like to 
understand the trade-offs between fidelity and probability for various photon detection goals.  In figures \ref{dakna_eta_fidelity_lambda}, 
\ref{dakna_eta_probability_lambda}, and \ref{dakna_eta_T_lambda} we plot the fidelity, ${\rm P}_{\eta}(m)$, and the beam splitter transitivity
$T$ as functions of $\lambda$ for $m=2$, $4$, and $6$.  In each case $T$ has been optimized to give the highest fidelity.  In each case, the maximum fidelity is achieved in the a region of low $\mathrm{P}_{\eta}(m)$.  The maximum fidelity in the presence of loss is equal to the maximum fidelity achievable when perfect photon counters are used.

We now expand our analysis to include two additional error sources: impurity in the squeezed state and ``dark counts'' in the photon counter.  Ideally, the squeezed state would be a pure state of a mode matched to the local oscillator.  However achieving this goal is extremely difficult, especially for pulsed squeezing experiments \cite{laporta1991,wasilewski2005,lvovsky2006,rohde2006}.  Instead, the squeezed state in the mode matched to the local oscillator is mixed, and other unwanted photons are traveling down the same beam path in other modes (for example, the wave packet carrying the extra photons has a different temporal profile).  These unwanted photons may be registered by the photon counter, falsely signaling the creation of a cat state.  For these reasons we expect that these unwanted photons will behave much like ``dark counts.''  The homodyne detection system will be immune to contamination by photons in non-matched modes, because homodyne detection is sensitive only to the mode exactly matched to the local oscillator.

\begin{figure}
\includegraphics[width=8.255cm]{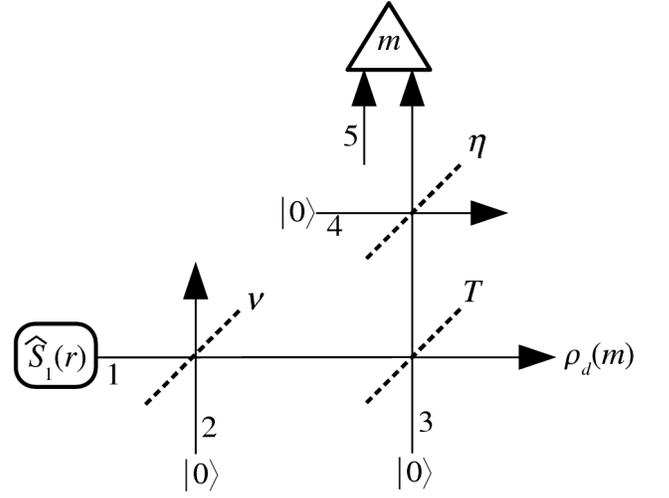}
\caption{Model of photon subtraction scheme including impurity in the initial squeezed state, inefficiency of the photon counter, and ``dark counts.''  The beam splitters with transmissivities $\nu$ and $\eta$ model loss from the squeezed state and inefficiency in the photon counter.  Modes 2 and 4 are lost to the environment, mode 5 contains ``dark counts'', and the output of mode 1 contains $\rho_d(m)$, which should be similar to a cat state.
\label{dakna_error_scheme}}
\end{figure}

We use the model in Fig.~\ref{dakna_error_scheme} to incorporate squeezed state impurity, photon counter inefficiency, and ``dark counts.''  The squeezing $\hat{S}_1(r)$ creates a pure squeezed state in mode 1.  Mode 1 then passes through a beam splitter with transmissivity $\nu$, and $1-\nu$ of the light is lost to the environment.  In any real squeezing experiment, the ideal squeezing and decoherence takes place simultaneously inside the down-conversion crystal.  This model of ideal squeezing followed by a beam splitter can create any state whose $x-$ and $p-$quadratures are Gaussian and centered at the origin, provided that the variance of one of the quadratures is equal to or less than the vacuum's variance.  The observed quadratures' variances are related to the squeezing $r$ and $\nu$ by
\be
v_x=\frac{1-\nu}{2}+\nu\frac{e^{-2r}}{2},
\ee
and
\be
v_p=\frac{1-\nu}{2}+\nu\frac{e^{2r}}{2}.
\ee
We use the conventions that the vacuum's variance is $1/2$ and $\hbar=1$.

The impure squeezed state in mode 1 then encounters the beam splitter with transmissivity $T$, which diverts light into mode 3.  To model the photon counter, mode 3 passes through the beam splitter, whose transmissivity $\eta$ is equal to the real photon counter's efficiency.  Mode 4 is lost to the environment.  In addition to mode 3, mode 5 also enters the photon counter.  The probability that mode 5 contains $q$ photons is given by the Poisson distribution with mean $d$, given in Eq.~(\ref{dark_count_probability}).  The actual ``dark count'' probability distribution may be different from the Poisson distribution, especially when many of the ``dark counts'' are actually caused by unwanted photons generated in the squeezing.

We calculate exactly the full state (that is, the $x$-quadrature wave functions) of the five modes symbolically with the Mathematica computer algebra software.  The output of mode 1, conditioned on $m$ photons being registered by the photon counter, is the mixed state $\rho_d(m)$.  We calculate the fidelity between $\rho_d(m)$ for typical values of $\nu$, $\eta$, and $d$.  We would like to optimize the fidelity by adjusting $r$, $T$, and $m$.  It is somewhat unrealistic to treat $r$, $\nu$, and $d$ as independently adjustable parameters, because they are all governed by dynamics occurring inside the down-conversion crystal.  However, this model should still be useful for making some general recommendations and predictions for experiments.

We consider two representative imagined experiments.  For the first, whose goal is to make an odd cat with $\alpha = 1.25$ by measuring $m=1$, we have $\nu=0.85$, $\eta=0.8$, $d=0.0005$.  The maximum achievable fidelity with these parameters is $F=0.727$ using $T=0.922$ and $r=-0.514$.  After the pure squeezed state passes through the beam splitter $\nu$, this corresponds to squeezed quadrature variances of 4.2 dB above the vacuum and -3.5 dB below.  The detector will register one photon (or one ``dark count'') with probability 0.0143.  

For the second experiment, the goal is to make an even cat with $\alpha = 2$ by measuring $m=2$.  For this more ambitious effort, $\nu=0.95$, $\eta=0.9$, and $d=0.0002$.  The best fidelity is now $F=0.737$ using $T=0.982$ and $r=-0.722$.  This corresponds to the observable squeezed quadrature variances of 6.11 dB above the vacuum and -5.62 dB below.  The detector will register two photons (or one photon and one ``dark count'' or two ``dark counts'') with probability 0.00021.

In the presence of significant levels of ``dark counts,'' the strategy of reducing $|r|$ and increasing $T$ to compensate for detector inefficiency is no longer effective.  There is now a maximum $m$ such that conditioning cat production on larger numbers of photons creates a lower fidelity cat state.  High purity of the initial squeezed state and matching of the squeezed state's spatial and temporal mode shape to the local oscillator is critical for making high fidelity cat states.  Any photons produced in the squeezing process that occupy modes other than that matched to the local oscillator will pollute the photon counter, degrading the cat fidelity.
It may be necessary to engineer the pump mode shape and the squeezing crystal to reduce the squeezing of other modes using methods similar to those described in \cite{grice2001,uren2005,raymer2005}.

In recent years some experiments have used photon subtraction to make small cat states, which are sometimes called ``Schr\"odinger kittens''.  All of these experiments used standard APDs to subtract the photons.  These detectors are unable to discriminate $1, 2, 3, \ldots$ photons, but the squeezing level is low, so the probability of more than one photon being present is very small.

The first of these experiments was performed by Wenger, Tualle-Brouri, and  Grangier \cite{wenger2004b}.  They used squeezed light of $r=-0.430$, from which they extracted a single photon using a beam splitter with $T=0.885$ to create an odd cat state.   They used a laser pulsed at 790 kHz and made 30,000 cat states in a three hour experiment.  This corresponds to a ${\rm P}_d(1) \sim 3.5 \times 10^{-6}$ or making 
$\sim 2.77$ cats/second.  They used tomography to reconstruct the Wigner function of this state, which clearly shows non-Gaussian characteristics.  However, the reconstructed Wigner function did not contain any negative values (a feature common to all cat states).  This is primarily because their experiment suffered from a high ``dark count'' rate, low photon detector efficiency, and impurity in the squeezed state.  A second pulsed experiment was performed by Orjoumtsev, Tualle-Brouri, Laurat, and Grangier \cite{ourjoumtsev2006}.  This improved on the previous experiment by having more careful filtering of the mode measured by the APD and higher purity in the initial squeezed state.  They produced cat states with $\alpha=0.89$, 0.84, and 0.78 having fidelities of 0.70, 0.64, and 0.63.

A continuous wave experiment was performed by Neergaard-Nielsen and co-authors \cite{neergaard-nielsen2006}.  Because light flows through their system continuously, but the photon counter gives discrete clicks, they recorded 1000 homodyne measurements lasting 2 $\mu$s before and after each APD click.  The series of homodyne measurements was then turned into a single quadrature value by weighting each data point according to a temporal mode function optimized to give the most negative Wigner function of the reconstructed state.  In this way they were able to effectively extract a single pulsed mode from the continuous wave.  They produced cat states with $\alpha=1.05$ and 1.3 with fidelities of 0.53.  A similar continuous wave photon subtraction experiment was performed by Wakui, Takahashi, Furusawa, and Sasaki \cite{wakui2006}.  They showed clear evidence of nonclassical states with negative Wigner functions, but they did not report cat state fidelities.

Progress in making higher fidelity and larger amplitude cats by photon subtraction can be achieved by carefully tuning the beam splitter transmissivity and the level of squeezing.  Until now, all experiments have subtracted only one photon from the squeezed state.  Using photon counters that can discriminate 0, 1, 2, ... photons would allow one to reject multi-photon events that degrade the fidelity of the cats produced in previous experiments.  Higher fidelities and amplitudes can also be achieved by subtracting more than one photon.  Subtracting two or more photons allows one to use a less squeezed state, which can be produced with higher purity.  However, the multimode nature of pulsed squeezed states may have a greater deleterious effect when more photons are subtracted.

\section{Making Kittens}
\begin{figure}
\includegraphics[width=8.255cm]{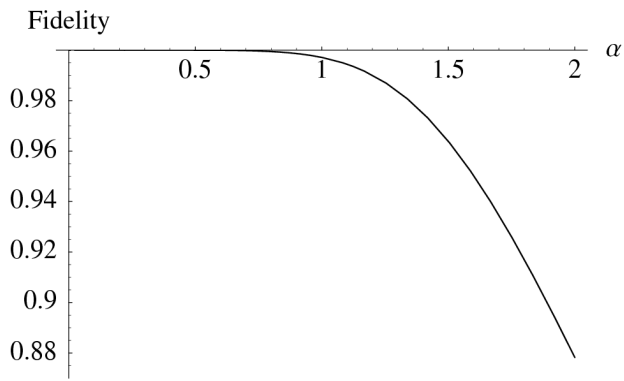}
\caption{Fidelity between an odd cat state of amplitude $\alpha$ and a squeezed single photon as a function of $\alpha$.
\label{fig:kittens_fidelity_vs_alpha}}
\end{figure}

\begin{figure}
\includegraphics[width=8.255cm]{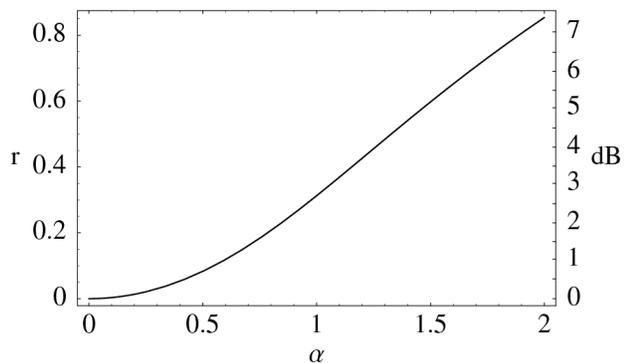}
\caption{Squeezing $r$ required to maximize the fidelity between a squeezed single photon and an odd cat of amplitude $\alpha$.
\label{fig:kittens_r_vs_alpha}}
\end{figure}

A small odd cat state is well approximated by a squeezed single photon~\cite{lund2004}. When the squeezing operator $\hat{S}(r)=e^{\frac{r}{2}
(\hat{a}^2-\hat{a}^{\dag2})}$ is applied to a single photon, the resulting state can be expanded in terms of photon number states as
\begin{equation}
\hat{S}(r)|1\rangle = \sum_{n=0}^{\infty}\frac{(\tanh r)^{n}\sqrt{(2n+1)!}}{(\cosh r)^{\frac{3}{2}}2^{n}n!}|2n+1\rangle,
\label{eq:squeezed_photon}
\end{equation} 
where $r$ is the squeezing parameter. The fidelity of this state with an odd cat state is given by
\begin{eqnarray}
F(r,\alpha)&=&|\langle\Psi_{-}(\alpha)|\hat{S}(r)|1\rangle|^2 
\nonumber \\ 
&=&\frac{2\alpha^2\text{exp}[\alpha^2(\tanh r-1)]}{(\cosh r)^3(1-\text{exp}[-2\alpha^2])}.
\label{eq:fid_sq_singphoton}
\end{eqnarray}
The maximized fidelity as a function of $\alpha$  is shown in Fig.~\ref{fig:kittens_fidelity_vs_alpha}\cite{lund2004}, and the $r$ that provides the maximum fidelity is shown in Fig.~\ref{fig:kittens_r_vs_alpha}\cite{kitten_attribution}.
The odd cat is well approximated by a squeezed single photon for small amplitudes $\alpha$ and small amounts of squeezing $r$.  A single photon squeezed by $r=0.31$ would have a fidelity of 0.997 with an odd cat of amplitude $\alpha=1$.

\begin{figure}
\includegraphics[width=8.255cm]{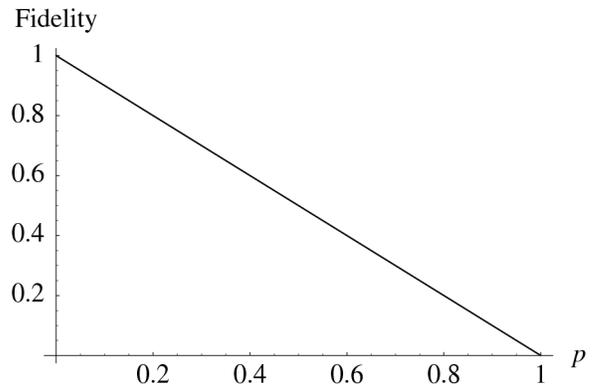}
\caption{Fidelity of the odd cat state with $\alpha=1/2$ with squeezed single photon state made from a photon source whose inefficiency is $p$.
\label{fig:CSSfid_sq_photon_vacuum}}
\end{figure}

There are numerous technologies for producing single photons.  Some of the most promising include heralded photons from nondegenerate down-conversion, optically and electrically excited quantum dots, and atoms and ions in optical cavities.  We refer the reader to \cite{singlephotons2004} for papers on single photon sources.  Unfortunately, pure single photon states are not produced using current technology. The single photon is always in a mixture with the vacuum as
\begin{equation}
p|0\rangle \langle0| + (1-p)|1\rangle \langle1|,
\label{eq:photon_mixed_vacuum}
\end{equation}
where $p$ is the inefficiency of the photon production. For most schemes, the probability to produce two or more photons is usually much smaller than $p$, so we ignore that possibility here. Therefore, the squeezed single photon state is also mixed with a squeezed vacuum
\begin{equation}
p\hat{S}(r)|0\rangle \langle 0|\hat{S}^{\dag}(r)+(1-p)\hat{S}(r)|1\rangle \langle1|\hat{S}^{\dag}(r).
\label{eq:squeezphoton_squeezvacuum}
\end{equation}
The squeezed vacuum is orthogonal to an odd cat state, so the fidelity of the state given by~(\ref{eq:squeezphoton_squeezvacuum}) and an odd cat state is $(1-p)F(r,\alpha)$.  We plot the fidelity of the state in Eq.~(\ref{eq:squeezphoton_squeezvacuum}) in Fig.~\ref{fig:CSSfid_sq_photon_vacuum} as a function of the photon production inefficiency $p$ for $\alpha=1/2$.

The success of this scheme to make kittens requires improvements in single photon production technologies.  It also requires a practical method to coordinate the single photon production with squeezing. One could attempt such an experiment by continuously pumping the down-conversion crystal, but homodyne measurements are recorded only when a single photon is (or is supposed to be) present.

\section{Growing Kittens Into Cats}

\begin{figure}
\includegraphics[width=8.225cm]{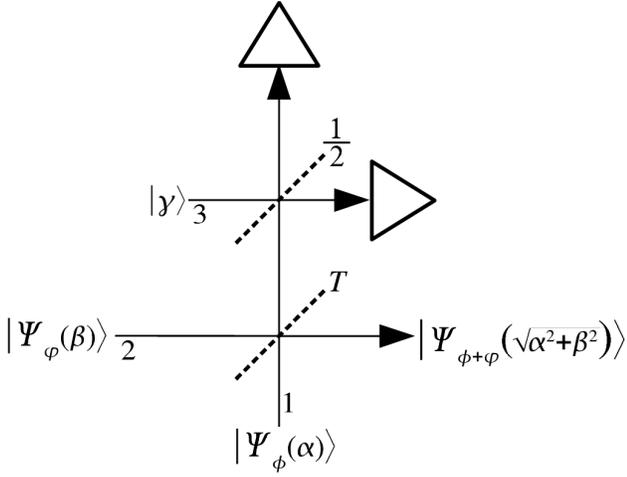}
\caption{Schematic of the nondeterministic amplification process to grow cat state amplitude.  Two small cat states at modes 1 and 2 and a coherent state in mode 3 are manipulated with linear optics.  A larger amplitude cat is produced in the output of mode 2 when photons are detected in both modes 1 and 3.
\label{fig:schemeCSS}}
\end{figure}

Lund, Jeong, Ralph, and Kim described a method for producing larger amplitude cat states given a source of kittens \cite{lund2004}.  This process uses only linear optical devices and photon detectors, and it is resilient to detector inefficiency.  The scheme is depicted in Fig.~\ref{fig:schemeCSS}.   

We begin with two small amplitude cat states $\ket{\Psi_\phi(\alpha)}_1$ and $\ket{\Psi_\varphi(\beta)}_2$ in modes 1 and 2.  The kittens meet in the first beam splitter, whose transmissivity is set to
\begin{equation}
T=\frac{\alpha^2}{\alpha^2+\beta^2}.
\label{eq:transmBS1}
\end{equation} 
One output of that beam splitter is sent to a second beam splitter, which has transmissivity $1/2$. Mode 3 contains the coherent state
\begin{equation}
\ket{\gamma}_3=\left|\frac{2\alpha \beta}{\sqrt{\alpha^2+\beta^2}}\right\rangle_3.
\label{eq:gamma}
\end{equation}
The two beam splitters transform the two small cats and the coherent state as
\begin{eqnarray}
&\hat{B}_{13}\left(\frac{1}{2}\right)\hat{B}_{12}\left(T
\right)\ket{\Psi_\phi(\alpha)}_1\ket{\Psi_\varphi(\beta)}_2|\gamma\rangle_3\propto 
\nonumber \\
&\left[\left|\frac{\gamma}{\sqrt{2}}
\right\rangle_1 \left(|-A\rangle_2 +e^{i(\varphi +\phi)}|A\rangle_2 \right)\left|\frac{\gamma}{\sqrt{2}}\right\rangle_3 \right. \nonumber \\
&+e^{i\phi}\left|\sqrt{2}\gamma\right\rangle_1 \left|\frac{\alpha^2-\beta^2}{A}\right\rangle_2|0\rangle_3\nonumber \\
&\left.+ e^{i\phi} |0\rangle_1 \left|-\frac{\alpha^2-\beta^2}{A}\right\rangle_2\left|\sqrt{2}\gamma\right\rangle_3\right],
\label{BS1BS2}
\end{eqnarray}
where $A=\sqrt{\alpha^2+\beta^2}$. When modes 1 and 3 exit from the second beam splitter, they are sent to photon detectors.  It is clear from Eq.~(\ref{BS1BS2}) that the resulting state for mode 2 when both detectors
register photons is a large cat $\propto (|-A\rangle_2 + e^{i(\varphi+\phi)} |A\rangle_2)$, whose amplitude $A=\sqrt{\alpha^2
+\beta^2}$ is larger than both $\alpha$ and $\beta$, and the relative phase is the sum of the relative phases of the input small cats.

\begin{figure}
\includegraphics[width=8.255cm]{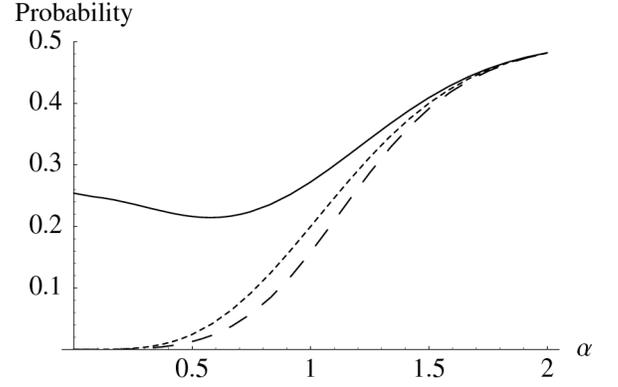}
\caption{Success probabilities for a single attempt at the process to make a cat from two kittens, depicted in Fig.~\ref{fig:schemeCSS}.  We show probabilities for the input fields of two identical small odd cat states (solid curve), two identical small even cats (dashed curve), and even and odd small cats (dotted curve).
\label{fig:probCSS}}
\end{figure}

The success probability (that both detectors detect at least one photon) for a single attempt of the process above is
\begin{equation}
\mathrm{P}_{\varphi,\phi}(\alpha,\beta)=\frac{(1-e^{-\frac{2\alpha^2\beta^2}{\alpha^2+\beta^2}})^2 [1+\text{cos}(\varphi+\phi)e^
{-2(\alpha^2+\beta^2)}]}
{2(1+\text{cos}(\varphi)e^{-2\alpha^2})(1+\text{cos}(\phi)e^{-2\beta^2})},
\label{eq:probCSS}
\end{equation}
which we plot for different input combinations in Fig.~\ref{fig:probCSS}\cite{lund2004}. The probabilities depend on the 
type of small cats (odd or even) used. The probability $P_{\pi,\pi}(\alpha,\alpha)$ for two identical odd cats is always larger than
$0.214$, regardless of the value of $\alpha$.

One could make the small cats necessary to produce a larger cat by this scheme, using any of the methods in this paper.

The photodetector's inefficiency will not affect the fidelity of the produced state.  It reduces only the success probability.  However, the detectors' dark counts may cause errors. Events in which the detector registers more photons than actually arrive at the detector can cause
us to accept a state that we should discard.  If the real number of photons coming to either of the detectors is zero, we should discard the resulting state, but dark counts can make the detector register a number of photons different
from zero and the resulting state would be accepted, decreasing the fidelity of the produced state. On the other hand, the inefficiency
of the detector decreases the probability of generating the correct state.  Here we provide new analysis of this scheme incorporating detector inefficiency and dark counts.

Although inefficiency alone would not affect the fidelity, to correctly model the effect of dark counts we must also include the photon detectors' inefficiency.  We imagine a beam splitter in front of each detector with transmissivity $\eta$.  We suppose that $l_1$ and $l_3$ photons pass through the beam splitters
before the detectors in modes 1 and 3.  This gives the state
\begin{eqnarray}
\rho_{\eta}(l_1,l_3)&=&\frac{1}{{\rm P}_{\eta}(l_1,l_3)}\sum_{n_1=l_1}^{\infty}\sum_{n_3=l_3}^{\infty}{\rm P}(n_1,n_3) \binom{n_1}{l_1} \binom{n_3}{l_3} \nonumber \\
 & & \times \eta^{l_1}(1-\eta)^{n_1-l_1} \times \eta^{l_3}(1-\eta)^{n_3-l_3} \nonumber \\
 & & \times |\psi(n_1,n_3)\rangle \langle \psi(n_1,n_3)|,
\label{eq:rho_eta_l1l2}
\end{eqnarray}
where $|\psi(n_1,n_3)\rangle$ is the normalized state given by the projection of Eq.~(\ref{BS1BS2}) into $\langle n_1,n_3|$, and ${\rm P}
(n_1,n_3)$
is the probability of measuring $n_1$ and $n_3$ photons in modes 1 and 3, respectively. To construct the density matrix of the
output mode after the dark counts we must add the $\rho_{\eta}(l_1,l_3)$ with their correct probability weights. This gives us
\begin{eqnarray}
\rho_d(m_1,m_3) & = & \frac{1}{{\rm P}_d(m_1,m_3)}\sum_{l_1=0}^{m_1}\sum_{l_3=0}^{m_3} {\rm P}_{\eta}(l_1,l_3)\times \nonumber \\
& & {\rm p}_d(m_1-l_1){\rm p}_d(m_3-l_3) \rho_{\eta}(l_1,l_3),
\label{eq:darkcounts}
\end{eqnarray}   
where ${\rm P}_{\eta}(l_1,l_3)$, ${\rm p}_d(m_1-l_1)$, and ${\rm P}_d(m_1,m_3)$ are given by Eqs.~(\ref{inefficiency_transformation}), (\ref{dark_count_probability}), and (\ref{dark_count_transformation}), respectively. $\rho_d(m_1,m_3)$ is the state produced when the detectors register $m_1$ and $m_3$ photons and dark counts.  The scheme is believed to have succeeded when both detectors register one or more photons, so the full state produced by this scheme is
\begin{equation}
\rho_{\mathrm{accept}} = \frac{\sum_{m_1=1}^{\infty} \sum_{m_3=1}^{\infty} {\rm P}_{d}(m_1,m_3) \rho_{d}(m_1,m_3)}{\sum_{m_1=1}^{\infty} \sum_{m_3=1}^
{\infty} {\rm P}_d(m_1,m_3)}.
\label{eq:rho_accept}
\end{equation}

The fidelity is $F=\langle \Psi_{\pm}(\alpha)|\rho_{\mathrm{accept}}|\Psi_{\pm}(\alpha)\rangle$. Let us consider the generation of 
a larger cat by this amplification scheme, given that we have two small odd cats as input. If we use
a photon counter whose efficiency is $\eta=0.8$ and an average number of dark counts per detection
event of $d=4 \times 10^{-4}$, we can generate an $\alpha=2$ cat with fidelity $F=0.9994$. If we use a TESPC, described in section \ref{Introduction}, whose 
efficiency is $\eta=0.88$ and the average number of dark counts per detection event is $d=10^{-8}$, we can obtain an $\alpha=2$ cat with a fidelity of $F=0.999999986$.  To obtain such an average number of dark counts we need to carefully filter out blackbody radiation, shield the experiment from 
ambient light, and restrict detection events to short time windows.

\begin{figure}
\includegraphics[width=8.255cm]{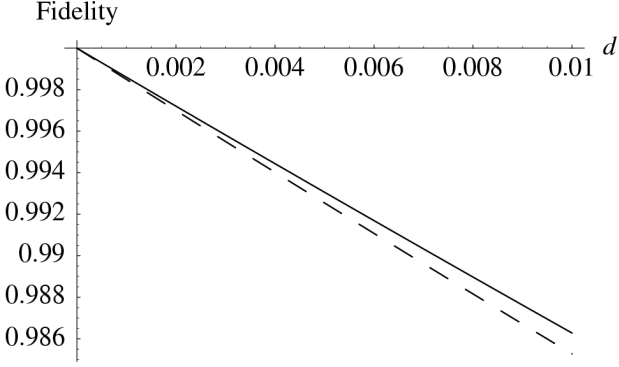}
\caption{Fidelity of a cat with $\alpha=2$ produced from kittens with $\alpha_i=\sqrt{2}$ as a function of the average number of dark counts in the detectors. The detectors have efficiencies $\eta=0.88$ on the solid line and $\eta=0.8$ on the dashed line.
\label{fig:fid_vs_d}}
\end{figure}

\begin{figure}
\includegraphics[width=8.255cm]{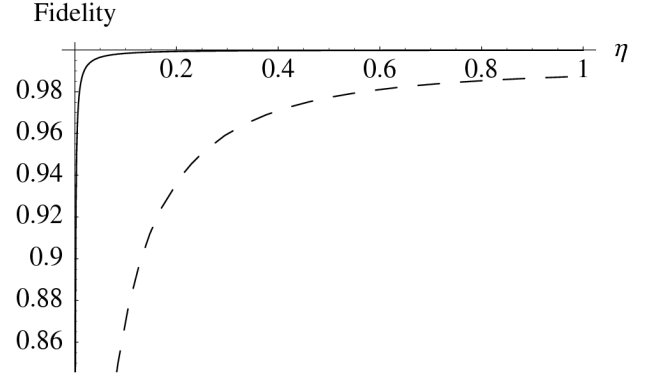}
\caption{Fidelity of a generated cat with $\alpha=2$ produced from kittens with $\alpha_i=\sqrt{2}$ as a function of the detectors' efficiency $\eta$. The detectors have mean dark counts $d=10^{-4}$ on the solid curve and $d=10^{-2}$ on the dashed curve.
\label{fig:fid_vs_eta}}
\end{figure}

Figs.~\ref{fig:fid_vs_d} and~\ref{fig:fid_vs_eta} show the fidelity as a function of the average number of dark counts and the efficiency, for generating an $\alpha=2$ cat state.

This scheme  has been analyzed in more detail by Jeong, Lund, and Ralph in \cite{jeong2004a}.  An important property they discussed is that, when using impure input kittens, this process can result in a purer output cat state.  That is, the output state can have a higher fidelity with a true cat than the fidelity of the input states.  Of course this is not generally true for any input state, but it is very effective when the input states are squeezed single photon kittens produced from inefficient single photon sources, as we discussed in the previous section.  This is because the cases in which a squeezed vacuum rather than a squeezed photon is input are more likely to be rejected because one of the photon detectors does not click.

%As shown in the last section, the squeezed single photon state can also
%be in a mixture with a squeezed vacuum, as given by Eq.~(\ref{eq:photon_mixed_vacuum}). The initial input states from the imperfect single photon source are
%\begin{eqnarray}
%\rho_{123}&=&\big[(1-p)^2 |S_1\rangle_1 {}_1\langle S_1|\otimes |S_1\rangle_2 {}_2\langle S_1| \nonumber \\
%& & + p^2 |S_0\rangle_1 {}_1\langle S_0| \otimes |S_0\rangle_2 {}_2\langle S_0| \nonumber \\
%& & + p(1-p)|S_0\rangle_1 {}_1 \langle S_0| \otimes |S_1\rangle_2 {}_2 \langle S_1| \nonumber \\
%& & + (1-p)p|S_1\rangle_1 {}_1\langle S_1| \otimes |S_0\rangle_2 {}_2\langle S_0| \big]\nonumber \\
% & & \otimes |\gamma\rangle_3 {}_3\langle \gamma|
%\end{eqnarray}
%where $|S_0\rangle = \hat{S}(r)|0\rangle$ and $|S_1\rangle = \hat{S}(r)|1\rangle$. The terms with $p^2$ and $p(1-p)$ are undesired error terms.

According to the calculations in \cite{jeong2004a}, if the probability that a single photon is not present is $p=0.4$ and the input squeezed single photons approximate cat states with amplitude $\alpha=1/2$, the fidelity of the initial small cat is $F=0.60$, but the fidelity of the output state after one iteration is $F=0.89$.  If $p=0.25$ $(p=0.05)$, the fidelity of the initial cat state is $F=0.750$ ($F=0.950)$ but becomes $F=0.941$ $(F=0.990)$ after one iteration of the cat growth procedure.

Jeong and co-authors also report that when one generates a large cat out of perfect squeezed single photon states, the fidelity maximizes for a particular number of iterations. Starting with smaller $\alpha_i$ states and performing more iterations to achieve $\alpha=2$ would produce a lower fidelity.  For example,
four iterations starting from the initial amplitude $\alpha_i=1/2$ are required to gain the maximum fidelity $F=0.995$ for $\alpha=2$.  High fidelity, $F>0.99$, can be obtained up to $\alpha=2.5$. 

A slight adaptation of the scheme in Fig.~\ref{fig:schemeCSS} was proposed by Rhode and Lund in \cite{rhode2007}.  They suggested using one even cat and one odd cat as input states.  The even and odd cats interfere at a beam splitter.  One of the output ports of the beam splitter is measured by a photon detector.  A larger amplitude odd cat state will appear in the other output port when the detector detects zero photons. Their paper also contains some analysis of the performance of this scheme in the presence of mode mismatch and loss.

Given a source of kittens, this appears to be an excellent scheme for producing larger amplitude cat states.  It can be implemented with current technology.  Its resilience to photon detector inefficiency and dark counts and its ability to purify its input are particularly attractive characteristics.

\section{Small Kerr Effect}

\begin{figure}
\includegraphics[width=8.255cm]{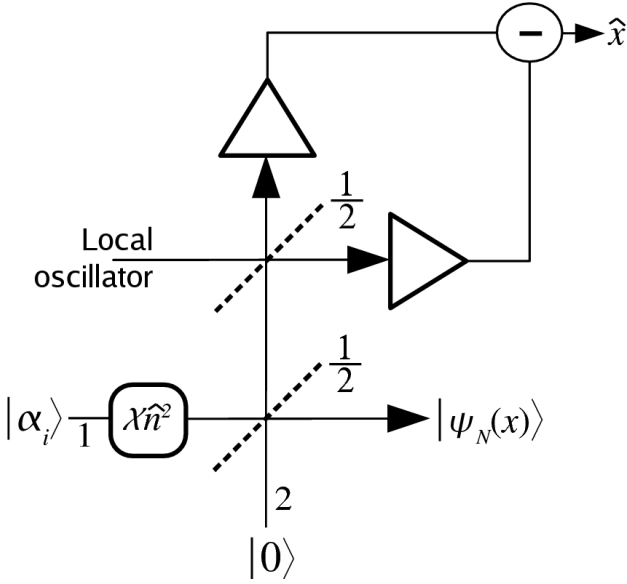}
\caption{Schematic diagram of the process to generate a cat state using small Kerr nonlinearity, a beam splitter, and a homodyne detection of the $\hat{x}$-quadrature.
\label{small_kerr_scheme}}
\end{figure}

In 2004, Jeong and co-authors \cite{jeong2004} proposed a probabilistic scheme to generate cat states using a small Kerr effect, a beam splitter, and a homodyne measurement.  This scheme uses a self-phase modulating Kerr medium like that discussed in Section \ref{section:kerr}, but the Kerr Hamiltonian is applied for a much shorter time, so photon loss should be a less significant factor.  A diagram of the scheme appears in Fig.~\ref{small_kerr_scheme}.

We begin the process by subjecting an initial coherent state $\ket{\alpha_i}_1$ in mode 1 to the Kerr Hamiltonian given by Eq.~(\ref{Kerr-Hamiltonian}). Under the influence of this Hamiltonian, an initial coherent state \mket{\al_{i}} will evolve to the following state after time $t$:
\be
|\psi\rangle_1 = e^{-\frac{|\alpha_{i}|^2}{2}} \sum_{n=0}^{\infty} \alpha_{i}^{n} \frac{e^{-i \phi_{n}}}
{\sqrt{n!}} |n\rangle_1,
\eel{eq:evolvedcoherent}
where $\phi_{n}=\chi t n^2$. The resulting state in Eq.~(\ref{eq:evolvedcoherent}) is an example of a generalized coherent state introduced by Titulaer and Glauber~\cite{titulaer1965} and 
discussed by Bialynicka-Birula~\cite{bialynicka-birula1968}. Generalized coherent states can always be
represented as superpositions of coherent states, with the same modulus but different phases. In the specific case of generalized coherent states for which
$\phi_{n}$ is periodic in $n$, they can be represented as a discrete superposition of $N$ coherent states.
For example, when the interaction time $\chi t$ is $\pi/N$ with a positive integer $N$, the initial coherent
state $\ket{\alpha}_1$ evolves to~\cite{lee1993}

\be
\ket{\psi_{N}}_1\propto\sum_{n=1}^{N} C_{n,N} \left|-\al_{i} e^{2in \pi /N}\right\rangle_1,
\eel{eq:psiN}
where we neglected the normalization. $C_{n,N}$ obeys 

\be
\sum_{n=1}^{N} C_{n,N}(-e^{2in\pi /N})^{k}=e^{-i \pi k^2/N}.
\eel{eq:coef1}
Using the method of Gantsog and Tana\'s~\cite{gantsog1991}, this can be solved to determine that

\be
C_{n,N}=\frac{1}{N} \sum_{k=0}^{N-1} (-1)^{k}\exp[-\frac{i \pi k}{N}(2n-k)].
\eel{ea:coef2}

The state in (\ref{eq:psiN}) is now combined with the vacuum state $\ket{0}_2$ in mode 2 on a 50-50 beam splitter. After passing through the beam splitter the system is in the state
\be
\sum_{n=1}^{N}C_{n,N}\left|-\al_{i}e^{2in\pi /N}/\sqrt{2}\right\rangle_1 \left|-\alpha_{i}e^{2in\pi /N}/\sqrt{2}\right\rangle_2.
\eel{eq:afterbeamspli}
We next perform a homodyne measurement of the $x$-quadrature of mode 2.  The measurement reduces the state of Eq.~(\ref{eq:afterbeamspli}) to
\be
\left|\psi_{N}(x)\right\rangle_1=\sum_{n=1}^{N}C_{n,N}(x)\left|-\alpha_{i}e^{2in\pi /N}/\sqrt{2}\right\rangle_1,
\eel{eq:nonlincat}
where
\be
C_{n,N}(x)=N_x C_{n,N} {}_2\left\langle x \bigg| -\alpha_{i}e^{2in\pi /N}/\sqrt{2}\right\rangle_2,
\eel{eq:coefnonlincat}
$N_x$ is a normalization factor, and $\ket{x}$ is the observed $x$-quadrature eigenstate.  To obtain the desired cat state, the measurement result is selected in such a way that the 
coefficients $|C_{N/2,N}(x)|$ and $|C_{N,N}(x)|$ have the same nonzero value and all the other coefficients are zero. For example, if $x=0$ is measured and $N=4k$, where $k$ is an integer,
the coefficients will be the largest for $n=N/4$ and $n=3N/4$. The coefficients will become smaller as $n$ gets
far from those two points.  

%A cat state could be obtained directly from~(\ref{eq:psiN}), if an interaction %time $\chi \tau$
%equals to $\pi/N$, with $N=2$ is considered. But to obtain such condition, with %typical values of $\chi$, 
%it is necessary to have a coherent state traveling a nonlinear cell of a length %where decoherence effects
%can not be neglected. 

The fidelity between the state $\left|\psi_N(x)\right\rangle$ and a perfect cat with appropriate amplitude is given by
\be
F(\al_{i},N,x) = \max_\phi\left[\left|\bra{\Psi_\phi(\alpha_i/\sqrt{2})} \psi_{N}(x)\rangle\right|^2\right].
\ee
Since the scalar product of two coherent states is
\begin{equation}
\langle\alpha|\beta\rangle=e^{-|\beta-\alpha|^2/2}e^{(\alpha^{*}\beta-\alpha \beta^{*})/2},
\end{equation}
we can write the fidelity as \cite{jeong2004}
\begin{eqnarray}
F & = & \max_\phi\left[N_\phi(\alpha_{i})^2 N_x^2\left|\sum_{n=1}^{N}C_{n,N}(x)\times \right.\right. \nonumber \\
 & & \left\{\exp\left[-\frac{\alpha_{i}^2}{2}\left(1+e^
{2in\pi/N}\right)\right] + \right.\nonumber \\
 & & \left.\left. \left. e^{i\phi} \exp\left[-\frac{\alpha_{i}^2}{2}\left(1-e^{2in\pi/N}\right)\right]\right\}\right|^2\right].
\label{eq:fidelity}
\end{eqnarray}

\begin{figure}
\includegraphics[width=8.255cm]{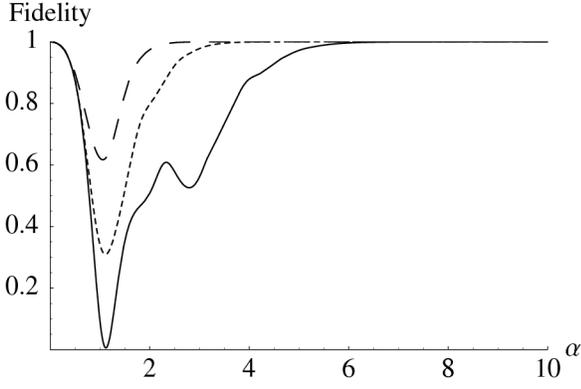}
\caption{Fidelity $F$ of the cat state produced by a weak Kerr effect when an $x=0$ measurement is obtained versus the generated cat amplitude $i \alpha$.  The solid curve shows the case when $N=20$ (one tenth of the interaction required by using the Kerr effect directly), the small dashes show $N=12$, and the long dashes show $N=8$.
\label{fig:fidvsalpha}}
\end{figure}

For an initial coherent state $\alpha_{i}$ and an interaction time $\chi t =\pi/20$ ($N=10$, one tenth of the interaction required to make a cat state using only the Kerr medium), the fidelity between
the state that leaves the Kerr medium and an ideal cat is $F \approx 0.1$ (for $\alpha_i=20$). After the homodyne measurement, the reduced state is similar to a cat state with amplitude $\alpha = i\alpha_{i}/ \sqrt{2}$. The maximum fidelity is achieved  when $x=0$ is observed. The output state is most similar to an even cat ($\phi=0$) state.  Figure \ref{fig:fidvsalpha} shows the fidelity of a generated cat when $x=0$ is measured as a function of the final amplitude $\alpha$. A good fidelity is obtained only for a final amplitude $\alpha$ larger than $\approx 5$ when $N=20$.

%\begin{figure}
%\begin{center}
%\includegraphics[width=80mm]{fid_only_kerr.eps}
%\caption{Fidelity $F$ of state that leaves the Kerr medium, after a interaction %time of $\lambda \tau =\pi/20$, as a function of the amplitude $\alpha$ }
%\label{fig:fidonlykerr}
%\end{center}
%\end{figure}

\begin{figure}
\includegraphics[width=8.255cm]{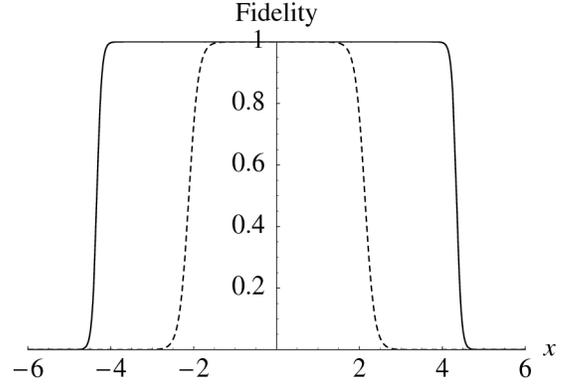}
\caption{Fidelity of the generated cat state against the measurement outcome $x$ for $N=20$.  The solid curve shows generated cat amplitude $\alpha=20i$, and the dashed curve shows $\alpha=10i$.  The fidelity has been maximized over the phase $\phi$ of the generated cat state.
\label{fig:fidvsX}}
\end{figure}

\begin{figure}
\includegraphics[width=8.255cm]{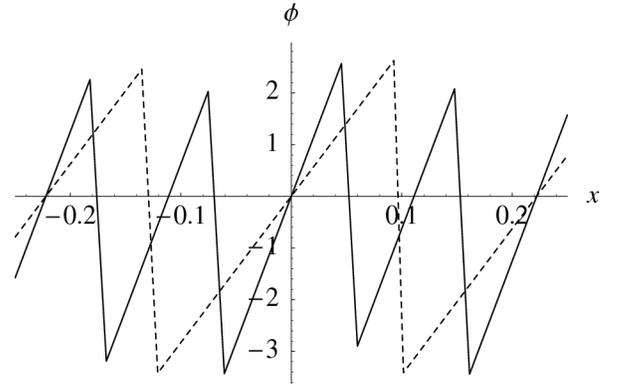}
\caption{Cat state phase $\phi$ that maximizes the fidelity shown in Fig.~\ref{fig:fidvsX}.  The solid curve shows generated cat amplitude $\alpha=20i$, and the dashed curve shows $\alpha=10i$.
\label{small_kerr_phi_vs_x}}
\end{figure}

The highest fidelity is achieved for the measurement result $x=0$.  We show a plot of the fidelity as a function of $x$ in Fig.~\ref{fig:fidvsX}.  The fidelity shown in the plot is the maximum fidelity for all possible values of $\phi$.  The phase of the cat state produced oscillates as a function of $x$, as shown in Fig.~\ref{small_kerr_phi_vs_x} \cite{small_kerr_attribution}.  Each cat's phase is random but can be inferred from the measurement of $x$.  Although the random phase is inconvenient, because it is known, one may correct or account for it.

To ensure high fidelity cats, one would choose some $\delta$ such that only measurement cases with $|x|<\delta$ are accepted.  The success probability to get an acceptably small $x$ is \cite{jeong2004}
\begin{eqnarray}
\mathrm{P}(\alpha_{i},N,\delta) & = & \int_{\delta} \mathrm{d}x \mathrm{Tr}_{12}\left[|\psi_n\rangle_{12}{}_{12}\langle\psi_n|\otimes|x\rangle_1 {}_1\langle x|\right]  
\nonumber \\
& =&\int_{\delta} \mathrm{d}x \sum_{n,m=1}^{N} \left\langle -\alpha_{i} e^{2in\pi/N}/\sqrt{2} \Big|x\right\rangle \nonumber \\
& & \times\left\langle x \Big|-\alpha_{i} e^{2im\pi/N}/\sqrt{2}\right\rangle \nonumber \\
& & \times \exp\left[-\frac{\alpha_{i}^2}{2}(1-e^{2i(m-n)\pi/N})\right],
\label{eq:probability}
\end{eqnarray}
where $|\delta|$ is the range of accepted $x$-quadrature measurements.  This ``success probability'' is the probability to obtain a cat of any phase $\phi$.  The probability to make a cat with a particular phase is much smaller.

One can make cat states with $\alpha=20i$ using $N=20$ and accepting all measurement results with $|x|<3.75$.  All of these states would have fidelities greater than 0.99997 with perfect cats with correctly chosen phases.  The probability to obtain $|x|<3.75$ is 0.052.  To make cats with $\alpha=10i$ and that have fidelities greater than 0.9991, one should accept $|x|<1.06$, which will happen with probability 0.045.  When making cat states with an even shorter interaction time (larger $N$), the fidelities are smaller and the success probability is lower.

%In Fig.~\ref{fig:fidvsint} we investigate the sensitivity of the production of %cat states to the precise value of the parameter
%$\chi \tau$. For that we find how the fidelity given by~(\ref{eq:fidelity}) %changes as the interaction time $\chi \tau$ changes.In
%this case, we have $\alpha_{i}=20$, what gives us a final cat with $\alpha=20$ %and $X=0$ measurement. We considered deviation 
%from the value $\chi \tau=\pi/20$ in a way that the new interaction time is %given by $\chi \tau=\pi/20-x$, where x is a percentage
%of the initial interaction time up to $1.5\%$.

%\begin{figure}
%\begin{center}
%\includegraphics[width=80mm]{fid-int-time.eps}
%\caption{Change in the fidelity F as a function of the change in the %interaction time $\chi \tau$. In this case $\alpha=20$,
%ideal interaction time $\chi \tau=\pi/20$, and $X=0$ is measured.}
%\label{fig:fidvsint}
%\end{center}
%\end{figure}

\begin{figure}
\includegraphics[width=8.255cm]{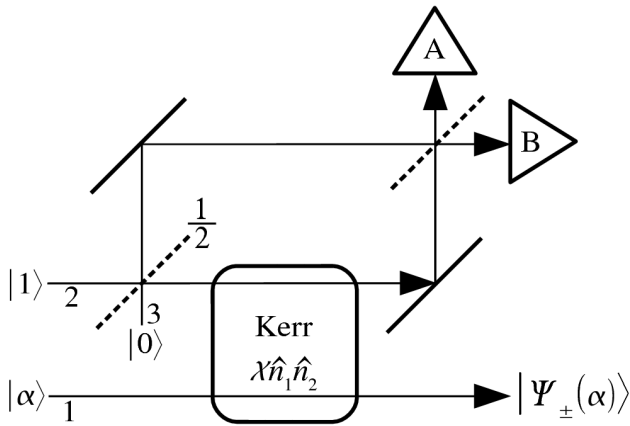}
\caption{Diagram of a scheme to produce cat states using a cross-phase modulating Kerr medium.  The dashed lines are beam splitters with transmissivity equal to $1/2$, and the solid bold lines are mirrors.  The triangles are photon detectors.
\label{fig:gerry}}
\end{figure}

A related scheme for using the Kerr effect was proposed by Gerry \cite{gerry1999}.  This scheme uses a cross-phase modulating type of Kerr medium, a single photon source, and photon detectors.  We picture the scheme in Fig.~\ref{fig:gerry}.

A single photon is inserted into a Mach-Zehnder interferometer, occupying modes 2 and 3.  The coherent state $\ket{\alpha}$ is in mode 1.  After the photon enters the interferometer, the three modes are in the state
\be
\frac{1}{\sqrt{2}}\left(\ket{\alpha,1,0}_{123}+\ket{\alpha,0,1}_{123}\right).
\ee
One arm of the interferometer (mode 2) and the coherent state in mode 1 pass through a Kerr medium, governed by the Hamiltonian $\chi \hat{n}_1 \hat{n}_2$.  After time $t$, this interaction evolves the system into the state
\be
\frac{1}{\sqrt{2}}\left(\ket{\alpha e^{-i\phi},1,0}_{123}+\ket{\alpha,0,1}_{123}\right),
\ee
where $\phi=\chi t$.  The phase of the coherent state is shifted when the photon accompanies it through the Kerr medium.  After modes 2 and 3 pass through the second beam splitter and exit the interferometer, the system is in the state
\begin{eqnarray}
\frac{1}{2}\left(\ket{\alpha e^{-i\phi},1,0}_{123}+\ket{\alpha e^{-i\phi},0,1}_{123}\right. \nonumber \\
-\ket{\alpha,1,0}_{123}+\ket{\alpha,0,1}_{123}\big).
\end{eqnarray}
When detector A sees 0 (1) photons and detector B sees 1 (0) photon, the resulting state of mode 1 is
\be
\frac{1}{\sqrt{2}}\left(\ket{\alpha e^{-i\phi}}_1\pm\ket{\alpha}_1\right).
\ee
When $\phi=\chi t=\pi$ this produces an even or odd cat with equal probability, depending on which of the two detectors registers the photon.

Like the method discussed in Section \ref{section:kerr}, this requires a Kerr medium with a large ratio of nonlinear strength to loss ($\chi/\gamma$) to make high fidelity cats.  However, in \cite{jeong2005} Jeong proposed an adaptation of this scheme designed to overcome the problem of loss in the Kerr medium.  Suppose one desires to make the cat state of amplitude $\alpha$.  Rather than inserting the coherent state $\ket{\alpha}$ into the network in Fig.~\ref{fig:gerry}, one should insert a coherent state $\ket{\beta}$ with a much larger amplitude $\beta \gg \alpha$.  One then applies the Kerr effect for a much shorter time, so that $|\beta e^{i\chi t}-\beta|=2\alpha$.  The displacement operator can shift the state $\ket{\beta e^{-i\chi t}}\pm\ket{\beta}$, so that it is symmetric about the origin of phase space.  This results in the desired state $\ket{\alpha}\pm\ket{-\alpha}$.

Jeong also analyzed the behavior when the Kerr medium absorbs photons.  If we increase $\beta$, the coherence of the desired output state, of fixed amplitude $\alpha$, also increases.  This is possible because one can subject the light to the lossy Kerr medium for a shorter time.  According to Jeong's estimates one could make a high fidelity cat with $\alpha=3$ using $\beta \sim 30,000$ through an optical fiber of  $~190$ m.  While Jeong's calculations show that this scheme is robust against loss in the Kerr medium, it requires displacing a superposition of two coherent states of many photons to the origin of phase space.  Small displacements can be achieved with high fidelity by passing the state through a beam splitter with transmissivity $T \rightarrow 1$, while a high amplitude coherent state passes through the other input of the beam splitter.  It is not clear how well this displacement scheme will work for states containing $\sim 30,000^2$ photons.  Any phase instability or mode mismatching in the displacement operation may lead to significant errors.

\section{Conclusions}

We have reviewed and analyzed several methods to make cat states.  Aside from making the cats directly using a strong, low loss Kerr effect, all of these methods involve preparation of the light using a lesser nonlinearity and postselecting on some measurement.  We attempt to compare some of the characteristics of the schemes in Table \ref{comparison_table}.

\begin{table*}
{\bf \caption{\label{comparison_table} Comparison of the Features of Cat Making Schemes }}
\begin{center}
\begin{tabular}{lp{4cm}cp{4cm}cp{4cm}}
\hline
                    & Detectors & \hspace{0.5cm} & Nonlinearity & \hspace{0.5cm} & Cat Produced\\
\hline
Direct Kerr effect & None & & Strong, low loss Kerr effect & & $F=1$ for zero loss \\
\hline
Back-action evasion & Efficient photon counting & & Squeezing twice & & High fidelity, low probability, low amplitude \\
\hline
Photon subtraction & Efficient photon counting & & Degenerate squeezing & & High fidelity, low probability, low amplitude \\
\hline
Squeezed single photon & None & & Single photon input, squeezing & & Very low amplitude \\
\hline
Kittens into cats & Photon detection low efficiency is OK & & None & & $F=1$ for perfect input kittens \\
\hline
Small Kerr effect & Homodyne & & Brief or weak Kerr effect & & High amplitude, fidelity conditional on homodyne measurement \\
\hline
\end{tabular}
\end{center}
\begin{flushleft}
\footnotesize
% put caption here
\normalsize
\end{flushleft}
\end{table*}

%\clearpage
%\begin{table}[h]
%{\bf \caption{\label{comparison_table} A comparison of the features of cat making schemes }}
%\begin{center}
%\begin{tabular}{|c|c|c|c|}
%\hline
%                    & Detector's efficiency & Dark counts & Squeezing \\
%\hline
%Back-action Evasion & Yes & No & Yes \\
%\hline
%Photon Subtraction & Yes & Yes &Yes \\
%\hline
%Kittens into Cats & No & Yes & No \\
%\hline
%Small Kerr Effect & Yes & No & No \\
%\hline
%\end{tabular}
%\end{center}
%\begin{flushleft}
%\footnotesize
%\normalsize
%\end{flushleft}
%\end{table}

We cannot make a definitive recommendation for the ``best'' cat making scheme for all circumstances.  This is because the schemes were formulated with different objectives, such as desired cat state amplitude, fidelity, and success probability; also the schemes each require significantly different resources and technologies to implement.  Nevertheless we can make a few comments on the relative utility of the schemes.

The photon subtraction scheme seems to be the simplest to implement for a laboratory demonstration of creating small amplitude ($\alpha \lesssim 2$) cats.  In fact a few experiments have already succeeded in this task \cite{ourjoumtsev2006,wakui2006}.  Improving on these experiments is likely to require making squeezed states with greater purity (especially in pulsed experiments) and subtraction of more than one photon.  Given the simplicity and fidelities achievable with the photon subtraction scheme, there seems to be little use in implementing any of the back-action evasion schemes.  The back action evasion schemes provide only slightly higher fidelities but require two stages of squeezing and are significantly more complex to implement.

If a large number of high fidelity small amplitude cats is required for use in a quantum computer, it may be advantageous to make the cats by squeezing single photons.  In this application it will be helpful to have the cats available on demand, which will require a high fidelity triggered single photon source, such as those based on quantum dots.  Quantum computation with cat states is likely to require fidelities significantly larger than 0.99, perhaps even 0.999 - 0.9999.  The iterative scheme for growing kittens to cats may serve to both increase cat amplitude and fidelity.  The combination of making kittens by squeezing single photons and then growing the kittens has the advantage that the resulting cat fidelity does not depend on the efficiency of photon detectors.

High amplitude cats may be most useful for interferometry applications.  In the future, making high amplitude cats might best be accomplished using the Kerr effect, either directly or after postselecting after homodyne detection as described in the ``Small Kerr Effect'' section.  Making high fidelity cats directly through the Kerr effect requires a loss to nonlinear strength ratio $\gamma/\chi \lesssim 1 $.  Single mode fused silica fibers have $\gamma/\chi\sim260$, and we have been unable to find any material with a smaller ratio.  Producing cats in this manner will likely require significant advances in technologies, such as EIT or photonic crystals.  Proposals for using a weaker Kerr effect followed by postselection may have less strict requirements for $\gamma/\chi$, but they are effective only when used with large numbers of photons, so large that the cats will be extremely susceptible to decoherence due to photon absorption and other errors.

Much previous research in Kerr materials has been devoted to increasing the interaction strength while suffering from much larger photon loss.  We strongly encourage research devoted to making low loss Kerr materials or interactions.  Low loss Kerr media will have many uses in addition to making cat states such as non-demolition photon counting and a controlled-phase quantum logic gate for single photon qubits.

%Any of these cat production methods can be combined with a scheme to use two small amplitude cats to make one larger amplitude cat.  This cat growth can be accomplished with beam splitters and inefficient photon detectors.  The cat growth scheme can also purify the cats when the input states are made from squeezing single photons from an inefficient source.

Mode mismatch is an error source that can affect any of the methods to make cat states.  While mode mismatch sometimes behaves similar to photon loss, its effects can be more subtle.  This is an issue that deserves more analysis than we have provided here.

\section*{Acknowledgments}
We thank Manny Knill for all of his helpful advice, and we thank Thomas Gerrits for help in preparing this manuscript. H. Vasconcelos thanks the Center for Applied Mathematics at University of Notre Dame for their financial support.  This paper is a contribution by the National Institute of
Standards and Technology and not subject to US copyright.  Certain trade names are identified in this report only in order to specify the methods used in obtaining the reported data. Mention of these products in no way constitutes endorsement of them. Other manufacturers may have products of equal or superior specifications. 

% \bibliography{journal_titles,mybib}

\end{document}